\documentclass[acmsmall,screen]{acmart}

\AtBeginDocument{%
  \providecommand\BibTeX{{%
    \normalfont B\kern-0.5em{\scshape i\kern-0.25em b}\kern-0.8em\TeX}}}

\copyrightyear{2020}

\setcopyright{acmlicensed}
\acmJournal{PACMHCI}
\acmYear{2020} \acmVolume{3} \acmNumber{CSCW} \acmArticle{TBD} \acmMonth{11} \acmPrice{15.00}\acmDOI{10.1145/3359134}

\usepackage[utf8]{inputenc}
\usepackage{enumitem}

\usepackage{wrapfig}

\usepackage{array,multirow,graphicx}
\newcommand{\STAB}[1]{\begin{tabular}{@{}c@{}}#1\end{tabular}}

\usepackage{pifont}
\newcommand{\cmark}{\ding{51}}%

\setcopyright{acmlicensed}

\acmDOI{0000001.0000001}

\received{June 2020}
\received[revised]{June 2020}
\received[accepted]{TBD}

\newcommand{\tableref}[1]{Table~\ref{#1}}

\newcommand\articlecount{4,238 }
\newcommand\revisioncount{973,940 }
\newcommand\editorcount{134,337 }

\begin{document}

\title{A Quantitative Portrait of Wikipedia's High-Tempo Collaborations during the 2020 Coronavirus Pandemic}

\author{Brian C. Keegan}
\orcid{0000-0002-7793-398X}
\affiliation{
  \institution{Department of Information Science, University of Colorado Boulder}
  \streetaddress{INFO 129, 1045 18th St.}
  \city{Boulder}
  \state{Colorado}
  \postcode{80309}
  \country{United States}
}
\email{brian.keegan@colorado.edu}

\author{Chenhao Tan}
\orcid{0000-0002-3981-2116}
\affiliation{
  \institution{Department of Computer Science, University of Colorado Boulder}
  \streetaddress{ECES 118A, 2055 Regent Drive}
  \city{Boulder}
  \state{Colorado}
  \postcode{80305}
  \country{United States}
}
\email{brian.keegan@colorado.edu}

\renewcommand{\shortauthors}{Brian C. Keegan \& Chenhao Tan}

\begin{abstract}
The 2020 coronavirus pandemic was a historic social disruption with significant consequences felt around the globe. Wikipedia is a freely-available, peer-produced encyclopedia with a remarkable ability to create and revise content following current events. Using \revisioncount revisions from \editorcount editors to \articlecount articles, this study examines the dynamics of the English Wikipedia's response to the coronavirus pandemic through the first five months of 2020 as a ``quantitative portrait'' describing the emergent collaborative behavior at three levels of analysis: article revision, editor contributions, and network dynamics. Across multiple data sources, quantitative methods, and levels of analysis, we find four consistent themes characterizing Wikipedia's unique large-scale, high-tempo, and temporary online collaborations: external events as drivers of activity, spillovers of activity, complex patterns of editor engagement, and the shadows of the future. 
In light of increasing concerns about online social platforms' abilities to govern the conduct and content of their users, we identify implications from Wikipedia's coronavirus collaborations for improving the resilience of socio-technical systems during a crisis.
\end{abstract}




\maketitle

\section{Introduction}\label{sec:introduction}
The coronavirus disease 2019 (COVID-19) pandemic was a historic social disruption with significant health, economic, political, and cultural consequences. Wikipedia is the ``free encyclopedia that anyone can edit'' with more than 6 million articles in English. Because Wikipedia's editors have a well-established and remarkable ability to create and revise encyclopedic content about current events~\cite{keegan_hot_2013,keegan_newswork_2013}, how did Wikipedia cover the pandemic? We offer a ``quantitative portrait'' characterizing Wikipedia's collaborations around topics related to the coronavirus pandemic focusing on article-level activity, user-level engagement, and network-level dynamics. 

The primary contribution of this paper is to provide theoretically-grounded 
descriptive insights into how volunteer editors working with twenty year-old interfaces and affordances 
are able to produce timely content about a historic global event. A deadly pandemic causing unprecedented social disruption around the globe seemingly presents a different scale of challenges to manage the coupling of diverse humans and distributed content artifacts in terms of the scope of content demanding editorial attention, the volatility and duration of information to process, and meeting shifting demands for information. 

This study examines the dynamics of the English Wikipedia's response to the coronavirus pandemic between December 2019 through May 2020. Using publicly-available digital trace data like revision history logs and archived article content for \revisioncount revisions by \editorcount editors to \articlecount articles related to the pandemic and its effects, we use mixed quantitative methods and inductive interpretations of these descriptive results to provide overlapping evidence around the structure and dynamics of an on-going and large-scale collaboration. We structure our analysis by focusing on analysis at the level of articles, editors, and networks. First, we analyze the changes in daily revision activity, similarity of article content, and pageview activity. Second, we examine the patterns of editors' engagement in these collaborations, the timing of their contributions, and the effects of their participation on their previous contributions. Finally, we use a social network perspective to understand how the structure of the collaboration network linking editors and articles together changes over time as well as how the structure of the hyperlink network of articles linking to each other has evolved. 

Drawing on perspectives from computer-supported cooperative work, organizational behavior, and crisis and public health informatics, we synthesize these descriptive quantitative findings through an inductive process to find four consistent themes around (1) external events as drivers of activity, (2) spillovers of activity from core to peripheral articles, (3) complex patterns of user engagement, and (4) the shadows of the future. These findings---while primarily descriptive and inductive in nature---characterize Wikipedia's unique large-scale, high-tempo, and temporary online collaborations with potential implications for designing more resilient socio-technical infrastructure. In the face of increasing concerns about online social platforms' ability to moderate incendiary content, intervene against disinformation campaigns, and limit mechanisms that drive polarization, Wikipedia's responses to current events serve as an important counterfactual about the pro-social, scalable, and resilient possibilities of online communication during a crisis.

\section{Background}\label{sec:background}
Wikipedia is a free online encyclopedia using a peer production model of volunteers authoring content as well as managing the community~\cite{jemielniak_common_2014,reagle_good_2010}. The production and consumption of information on Wikipedia is a matter of theoretical, empirical, and popular interest given its prominence as one of the most visited websites while also being a volunteer-led online social platform. Wikipedia's editors, their managerial systems, and their reliance on technological tools makes it a canonical example of a \textit{socio-technical system}~\cite{niederer_wisdom_2010}. Since its earliest days, Wikipedia's online community of editors have been able to use this platform to rapidly create and update encyclopedic content in the aftermath of accidents, disasters, and other crises. The communities of Wikipedia volunteers who rapidly and temporarily self-organize to author and edit content about breaking news and current events are examples of \textit{high-tempo collaborations}. This section reviews prior work on 
coordination, crisis informatics and digital epidemiology, and Wikipedia's high-tempo collaborations to motivate our analysis of Wikipedia's coverage of the 2020 coronavirus pandemic.


\subsection{High-tempo, high-reliablity, and temporary coordination}
Coordination is the management of dependencies between activities to accomplish tasks~\cite{malone_interdisciplinary_1994,okhuysen_coordination_2009}. Coordination is a particularly important construct for understanding the processes of sharing, transfer, accumulation, transformation, and co-creation in knowledge collaboration~\cite{faraj_knowledge_2011}. Classic approaches to coordination emphasize task uncertainty, task interdependence, and work unit size as crucial determinants of impersonal (rule and schedules), personal (horizontal or vertical communication), or group  (meetings) coordination modes~\cite{van_de_ven_determinants_1976}. In contrast to these \textit{explicit} coordination mechanisms, teams can also employ \textit{implicit} coordination mechanisms that rely on anticipation~\cite{wittenbaum_tacit_1996}, heedfulness~\cite{weick_collective_1993}, shared mental models~\cite{cannonbowers_shared_1993}, and other informal strategies~\cite{rico_implicit_2008}. But many collaborative contexts---particularly online---lack formal leadership, defined roles, and predictable work units for explicit coordination and the social information like identity, familiarity, and expectations required for implicit coordination. 

The teams typically studied by organizational researchers rarely face the unpredictability, urgent demands, and immediate consequences as groups in \textit{high-tempo} contexts like disaster response~\cite{majchrzak_coordinating_2007}, emergency medicine~\cite{faraj_coordination_2006}, carrier flight decks~\cite{weick_collective_1993}, SWAT teams~\cite{bechky_expecting_2011}, and journalism~\cite{berkowitz_nonroutine_1992}. \textit{Temporary organizations} are where people work together on complex, non-routine, consequential, and high-risk tasks and disband after a deadline or resources are expended~\cite{bakker_taking_2010,bakker_temporary_2016,burke_temporary_2016}. \textit{High-reliability organizations} operate in technologically complex settings like nuclear reactors, aircraft cockpits, and trading desk where errors can quickly cascade into into catastrophic consequences so there is an overwhelming focus on safety ~\cite{bigley_incident_2001,weick_organizing_1999,dekker_high_2010}.
These frameworks all characterize distinct types of organizations with varying degrees of overlap: temporary organizations do not need to be high-tempo or high-reliability (a board game night), high-tempo organizations do not need to be temporary (a carrier flight deck), high-reliability organizations do not need to be high-tempo (a nuclear reactor control room), and temporary organizations can still be high-reliability (an emergency room). Wikipedia's breaking news collaborations can be characterized as \textit{high-tempo} because of the rapid pace of editing activity from dozens or hundreds of users in response to new information about an event, \textit{high-reliability} because contributions are immediately live to an audience of thousands of information-seeking users, and \textit{temporary} because eventually information about an event saturates and editors' interests shift to other topics.

However, many organizations sharing these characteristics employ similar coordination strategies: improvisation~\cite{barrett_improvisation_1998}, regeneration~\cite{bechky_gaffers_2006},  jamming~\cite{eisenberg_jamming_1990}, representation~\cite{kellogg_marketing_2006}, delegation~\cite{klein_delegation_2006}, heedful interrelating~\cite{weick_collective_1993}, swift trust~\cite{meyerson_swift_1996}, collective responsibility~\cite{valentine_scaffolds_2014}, emergent roles~\cite{arazy_emergent_2016}, reciprocal exchange~\cite{faraj_exchange_2010}, sensitivity to operations~\cite{weick_managing_2007}, mindfulness~\cite{sutcliffe_mindfulness_2016}, privileging expertise~\cite{faraj_expertise_2000}, rendering work visible~\cite{kellogg_marketing_2006}, vigilant interactions~\cite{jarvenpaa_vigilant_2010}, interdependency awareness~\cite{benmenahem_coordinating_2016}, folding~\cite{shaikh_folding_2016}, and knowledge shaping~\cite{yates_shaping_2010}. It is beyond the scope of this study to summarize or synthesize the specifics of all these mechanisms, but to instead emphasize that many kinds of high-tempo, high-reliability, and/or temporary organizations have developed mechanisms for coordinating under stress that could be implemented, adapted, or re-discovered by Wikipedia editors to support their breaking news collaborations. Some Wikipedia-specific mechanisms are summarized in a later section of this background.

\subsection{Crisis informatics and digital epidemiology}
The emergence of self-organizing altruistic communities engaged in recovery work in the aftermath of crises is one of the most robust findings in disaster sociology~\cite{drabek_emergent_2003,quarantelli_response_1977,stallings_emergent_1985}. With the proliferation of wireless, mobile, and social technologies, people use digital communication to respond to disaster warnings, responses, and recovery~\cite{palen_crisis_2016}. The field of crisis informatics has primarily focused on the use of social media to understand the dynamics of emergent communities during disasters~\cite{palen_emergence_2008,shklovski_finding_2008}, but other information technologies like SMS messages~\cite{palen_citizen_2007}, chatrooms~\cite{norris_is_2019}, and geographic information systems~\cite{soden_infrastructure_2016} are also active sites of technologically-mediated post-disaster engagement. The ``improvisation of order out of chaos'' alongside collective feelings of equanimity and egalitarianism~\cite{solnit_paradise_2009}, while temporary, align well with many of Wikipedia's core values like neutrality and openness~\cite{jemielniak_common_2014,reagle_good_2010}. However, Wikipedia is generally not a primary data source for crisis informatics researchers or emergency managers because of the lack of geo-location and lags in reporting pageview data.

Digital epidemiology and public health informatics use digital trace data like search engine query logs and social media posts to augment and accelerate classic methods for surveilling for diseases~\cite{salathe_digital_2012,santillana_combining_2015,althouse_enhancing_2015,ginsberg_detecting_2009,sharpe_evaluating_2016}. Wikipedia is a prominent online resource for health information with adequate quality information for consumers~\cite{heilman_wikipedia_2015,shafee_evolution_2017,smith_situating_2020}. Wikipedia's WikiProject Medicine has distinguished itself through concerted efforts within the community and outreach to medical professions and researchers to expanding and improving the quality of health information~\cite{heilman_wikipedia_2011,kane_content_2016,ransbotham_network_2012}. Because Wikipedia consistently ranks in the first ten results of many health queries~\cite{laurent_seeking_2009} and it publicly shares access log data in the form of pageviews, a growing body of literature explores Wikipedia's potential as an epidemiological data source. Specifically, pageviews to Wikipedia articles about diseases reliably burst during an outbreak as people seek information about the disease in response to media coverage~\cite{shafee_evolution_2017,al_tamime_observing_2018,tausczik_public_2012,geis_interplay_2016}. While some studies emphasize the feasibility of using Wikipedia data to forecast influenza~\cite{generous_global_2014,mciver_wikipedia_2014,tausczik_public_2012,zimmer_use_2018,hickmann_forecasting_2015,bardak_prediction_2015}, dengue~\cite{heilman_dengue_2014}, zika~\cite{tizzoni_impact_2020}, and other diseases~\cite{fairchild_eliciting_2015,tamime_uncertainty_2019}, other research concludes behavioral heterogeneity on social platforms like Wikipedia cannot reliably beat canonical epidemiological forecasting methods~\cite{priedhorsky_measuring_2017,althouse_enhancing_2015,osthus_even_2019,lazer_parable_2014}. 

We emphasize the goal of our study is \textit{not} to evaluate the effectiveness of Wikipedia data in supporting emergency management or forecasting coronavirus incidence. Instead, we expect that coronavirus-related activity on Wikipedia will mirror previous disease outbreaks in terms of spikes of information seeking and contributions from editors of medical topics. Preliminary work has also examined shifts in collective attention on Wikipedia as a result of quarantining efforts~\cite{ribeiro_sudden_2020} and citation practices on COVID-19 related articles~\cite{colavizza_covid19_2020}. Because the effects of the pandemic are greater than any outbreak since Wikipedia's founding in 2001~\cite{harrison_coronavirus_2020,qaiser_like_2020}, we expect the scale of information production and seeking will demand new forms of collaboration and coordination that have not been seen before. 

\subsection{High-tempo collaborations on Wikipedia}

\begin{figure}
    \centering
    \includegraphics[width=.9\textwidth]{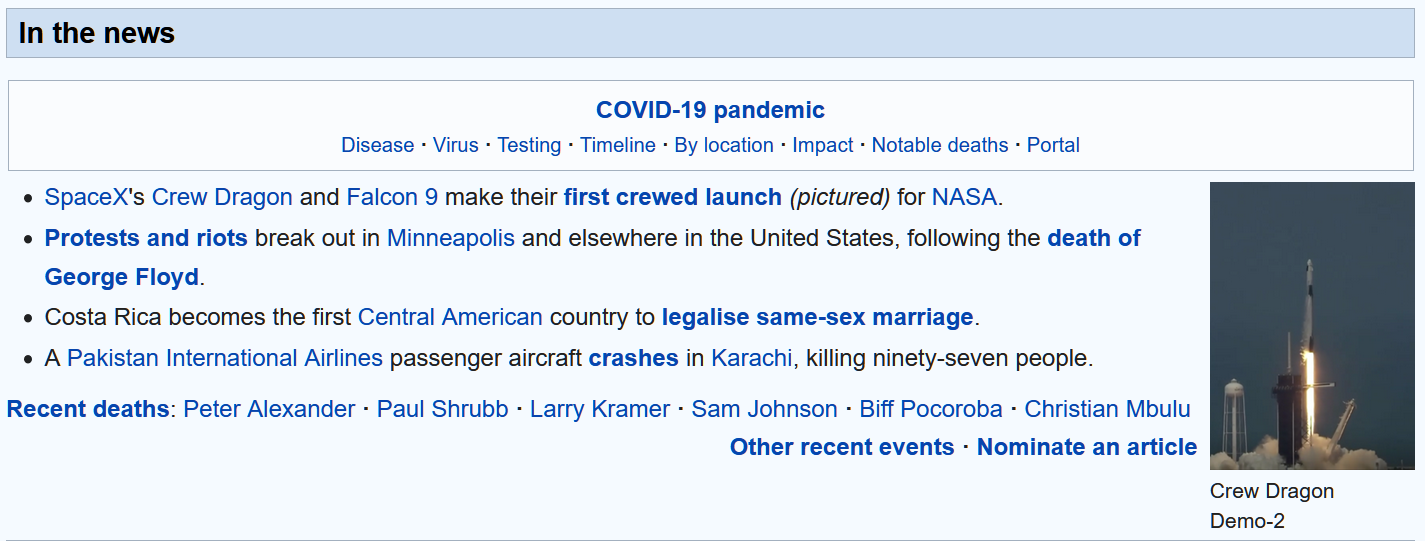}
    \caption{An example of the ``In the News'' (ITN) template on Wikipedia's homepage featuring articles related to current events. During the pandemic, the ITN template had a dedicated ``special header'' devoted to COVID-19 related topics (at top) like the pandemic, disease, virus, testing, local responses, timeline, and deaths.}
    \label{fig:itn}
\end{figure}

Despite early apprehensions about its radical ``anyone can edit'' model polluting the web with misinformation~\cite{denning_wikipedia_2005,messner_legitimizing_2011}, Wikipedia's responsiveness to current events is a major factor in its identity, rules, and history~\cite{keegan_newswork_2013}. The creation and revision of Wikipedia content in the aftermath of major events is not done automatically or through formal delegation, but is accomplished by high-tempo collaborations in which large numbers of volunteers---most of whom have never worked together before---self-organize into intense and temporary collaborations using twenty-year-old interfaces and affordances. These collaborations are an important boundary case that challenges our assumptions and theories about whether online communities can succeed~\cite{bryant_becoming_2005,kraut_building_2012} in the face of unstable membership, asynchronous interaction, diverse motivations, unclear leadership, 
and volatile information. But the quality of the work done in these collaborations is reliably excellent: Wikipedia articles about current events have been cited as exemplars of timeliness, breath, and accuracy around events like U.S. presidential elections~\cite{boxer_mudslinging_2004,cohen_dont_2008}, school shootings~\cite{cohen_latest_2007}, earthquakes~\cite{montgomery_minutes_2011}, and other events~\cite{dee_all_2007,flowers_edited_2016}. The risks and opportunities associated with Wikipedia's coverage of current events have likewise been noted by countries banning access~\cite{harrison_how_2020,siegel_search_2019}, politicians polishing their articles~\cite{estes_wikipedia_2011,gobel_political_2018}, and corporations spinning controversies~\cite{hafner_seeing_2007,etter_collective_2015}.

Like other forms of high-reliability organizing, overlapping practices and resources enable Wiki\-pedia editors to engage in these unusual collaborations. Editors delegate labor-intensive tasks using bots and templates~\cite{geiger_when_2013,keegan_hot_2011,ford_infoboxes_2015}, regenerate structures from previous crises~\cite{keegan_hot_2013}, alter their interaction practices~\cite{keegan_staying_2012}, adopt emergent social roles based on their previous experience~\cite{keegan_emergent_2015}, develop routines to handle unpredictable but regular events~\cite{keegan_posthumous_2015}, migrate between articles to ensure their consistency~\cite{twyman_black_2017}, respond to media coverage and agenda setting~\cite{keegan_presidential_2019}, and adding culturally-salient media~\cite{porter_visual_2020}. Wikipedia is a site of collective memory formation for major historical events like the Vietnam War~\cite{luyt_wikipedia_2016}, September 11 attacks~\cite{ferron_beyond_2014}, Arab Spring~\cite{ferron_collective_2011,ferron_wikirevolutions_2011}, and deaths of celebrities~\cite{keegan_posthumous_2015}. Because Wikipedia preserves the content of every editor's revision to its articles, it is possible to mine these histories to observe the evolution of framings and interpretations around individuals, institutions, and events~\cite{pentzold_fixing_2009,pentzold_digging_2017,garciagavilanes_memory_2017,kanhabua_what_2014}. Collective memory repertoires like re-appraising historical events during a contemporary event can be significant drivers of attention and information demand~\cite{twyman_black_2017}. The evolution of Wikipedia's content around the coronavirus will be of immense interest to future scholars to explore how our understanding of the virus and its implications changed or how Wikipedians' discussions reflected larger debates~\cite{harrison_future_2020}.


\section{Approach}\label{sec:approach}

In this section, we discuss the details of our data collection and sampling procedures.

\subsection{Sampling}

\subsubsection{Articles}
The English Wikipedia has standalone articles for the \textit{event} (``\href{https://en.wikipedia.org/wiki/COVID-19_pandemic}{COVID-19 pandemic}''), the \textit{disease} (``\href{https://en.wikipedia.org/wiki/Coronavirus_disease_2019}{Coronavirus disease 2019}''), and the \textit{virus} (``\href{https://en.wikipedia.org/wiki/Severe_acute_respiratory_syndrome_coronavirus_2}{Severe acute respiratory syndrome coronovirus 2}''). We focus on the event article---``COVID-19 pandemic''---
as the \textit{seed article} for subsequent sampling steps because it was the largest, longest, most active, and more viewed of the three primary articles. We retrieved related articles through an iterative snowball sampling approach to identify a set of related articles. This iterative sampling methodology consisted of several steps. First, we extracted all the member and child pages of the primary category (i.e., ``Category:COVID-19''), retrieving 3,069 pages in English. Examples of these \textit{category articles} include ``2019–20 coronavirus outbreak by country and territory'', ``Severe acute respiratory syndrome coronavirus 2'', and ``Template:2019–20 coronavirus outbreak data''. Second, we identified prior revisions of the pandemic event article at a weekly frequency going back to Sunday, 5 January 2020. For each weekly revision of the seed article, we retrieved the hyperlinks from that week's version of the article to \textit{neighbor articles}. We combined these weekly neighbor articles together into a super-set of neighboring articles that were ever linked to from the seed article over these revisions.

The first two steps give us two distinct sets of similar articles: \textit{category-related articles} and \textit{neighbor-related articles}. The articles related by category (the first step), by definition, have extremely high topical similarity but also include administrative pages like templates. The articles related by hyperlink neighbors (the second step) are related to the event in some way, ranging from obvious relationships (\textit{e.g.}, ``Wuhan'') to less obvious relationships (\textit{e.g.}, ``Diabetes''). Combining both of these sampling approaches enables us to balance precision (category-related articles) and recall (neighbor-related articles) for a set of articles likely impacted by the coronavirus pandemic. We use the union of both of these samples as the set of articles in the rest of the analysis: 717,776 revisions to 3,064 category-related articles and 256,174 revisions to 1,174 neighbor-related articles for a total of \revisioncount revisions to \articlecount articles from \editorcount editors.
We note this is not the only sampling approach possible to retrieve COVID-19 related articles: the Wikimedia Foundation publishes summary statistics of revisions and pageviews over a set of related pages derived from Wikidata relationships.\footnote{\url{https://covid-data.wmflabs.org/}} \tableref{tab:articles} shows example pages in each type of Wikipedia articles that we consider in this work.

\subsubsection{Editors.} The third step in our sampling approach is to identify a relevant set of users contributing to these articles. Because user-level revision activity is highly right-skewed and distributed across thousands of articles with their own local editing microclimates, we instead prioritize retrieving the contribution histories for all the registered editors who contributed to the event, disease, and virus articles from December 2018 through May 2020.


\subsection{Information retrieval} 
For each article in the union of category-related and neighbor-related articles and the contributors, we retrieved data from December 2018 through May 2020. This sampling window captured a baseline of revision activity over an approximately 12-month period before the outbreak and pandemic as a baseline. We used a custom and openly-licensed Python library\footnote{Link to library removed for anonymization.} to retrieve data from the official Wikipedia API endpoints about revision histories, user contributions, revision content, and pageviews. 
\begin{description}[leftmargin=1em]
    \item[Page revision histories.] The revision histories of each of the \articlecount articles were retrieved and include the revision ID, the name or IP address of the user who committed the revision, the timestamp of the revision, the size of the page at the time, and a SHA-1 hash of the page content to aid in revert tracking. Secondary features like the size of the changes introduced by the revision (``diff''), the time elapsed since the previous revision (``lag''), and the time elapsed since the page was created (``age'') were also competed for each revision.
    \item[Editor contribution histories.] The contribution histories of a sample of users who contributed to articles was also retrieved. 
    While this is only 3\% of the users across all the articles, it still captures 178,550 revisions, or 20\% of all observed revisions in our sample. 
    These users' contribution histories across \textit{all} articles and other pages (in all namespaces) were retrieved from December 2018 through May 2020, generating 3.9 million unique revisions. Like the revision history data, these data include the revision ID, the timestamp, namespace of the page, and size of the revision. Secondary features like the size of the changes and time elapsed since the previous revision were also computed.
    \item[Revision content.] The content of individual historical revisions were retrieved from the API. The full HTML of these revisions were stored and custom scripts parsed them into structured data like lists of hyperlinks, external links, references, and plain text for analysis over time. These revisions were only sampled on a weekly frequency, retrieving the last revision of each article every Sunday.
    \item[Pageviews.] The number of times a page was served through a website or mobile device was retrieved from the official API. These pageview data are available at a daily frequency and are not disaggregated by geography, language, \textit{etc}. in order to preserve privacy. We retrieve the ``all-access user'' data combining web and mobile views but excluding views from agents likely to be spiders/bots. Pageview data is assigned to the article even if the article is a redirect~\cite{hill_consider_2014}, so the redirect graph for each article in our sample was constructed, the pageviews for each of those primary and redirected articles retrieved, and the pageview numbers for the focal and redirect articles were aggregated together.
\end{description}

\begin{table}[t]
    \footnotesize
    \centering
    \begin{tabular}{clc}
        \toprule
        \textbf{Label} & \textbf{Event(s)} & \textbf{Date(s)} \\ \midrule
        A & China reports first case & 2 January 2020  \\
        B & China begins quarantines, U.S. reports first cases & 22--25 January 2020 \\ 
        C & \textit{Diamond Princess} quarantined in Japan & 4 February 2020 \\
        D & U.S. reports community transmission and first death & 26-29 February 2020 \\
        E & Italy quarantines; schools and pro-sports canceled; ``Black Thursday'' market crash & 6--13 March 2020 \\
        F & Statewide stay-at-home orders begin & 16--23 March 2020 \\
        \bottomrule
    \end{tabular}
    \caption{U.S.-centered timeline of events}
    \label{tab:timeline}
\end{table}


\section{Article-level Activity}\label{sec:article}
Using the articles as a primary unit of analysis, we examine how the activity related to revisions, pageviews, and content similarity changed for the three seed articles (the pandemic {\em event}, the {\em disease}, and the {\em virus}) as well as the larger sample of category-related and neighbor-related articles. 

\begin{table}[t]
    \footnotesize
    \centering
    \begin{tabular}{lcccccc}
        \toprule
        \textbf{Article title} & \textbf{Type} & \textbf{Created} & \textbf{Revisions} & \textbf{Editors} & \textbf{Pageviews} & \textbf{Size} \\ \midrule
        COVID-19 pandemic & Seed & 2020-01-05 & 19,858 & 2,749 & 63,833,280 & 332,803 \\
        Coronavirus disease 2019 & Seed & 2020-02-05 & 4,255 & 881 & 14,486,983 & 230,522 \\
        SARS coronovirus 2 & Seed & 2020-01-09 & 2,919 & 646 & 7,374,993 & 79,997  \\ \midrule
        Template:COVID-19 pandemic data & Category & 2020-01-28 & 24,083 & 1,001 & 269,827 & 171,627 \\
        2020 Democratic Party primaries & Category & --- & 8,628 & 1,417 & 9,494,957 & 274,742 \\
        COVID-19 pandemic in the U.S. & Category & 2020-02-17 & 7,730 & 1,330 & 12,921,675 & 382,169 \\ \midrule
        2019--20 Hong Kong protests & Neighbor & 2019-06-10 & 7,383 & 1,156 & 432,714 & 443,948 \\
        Donald Trump & Neighbor & --- & 3,941 & 456 & 9,989,039 & 400,926 \\
        Boris Johnson & Neighbor & --- & 2,074 & 650 & 7,051,692 & 253,617 \\
        \bottomrule
    \end{tabular}
    \caption{Basic information for the seed, category-related, and neighbor-related English Wikipedia articles for activity between December 2018 and May 2020. 
    We show three examples of category-related and neighbor-related articles.
    Some article titles have been abbreviated for space. Page creation dates for articles that were created before December 2018 are not reported.
    }
    \label{tab:articles}
\end{table}

\subsection{Revisions}

\begin{figure}[t]
    \begin{minipage}{.475\textwidth}
        \centering
        \includegraphics[width=\textwidth]{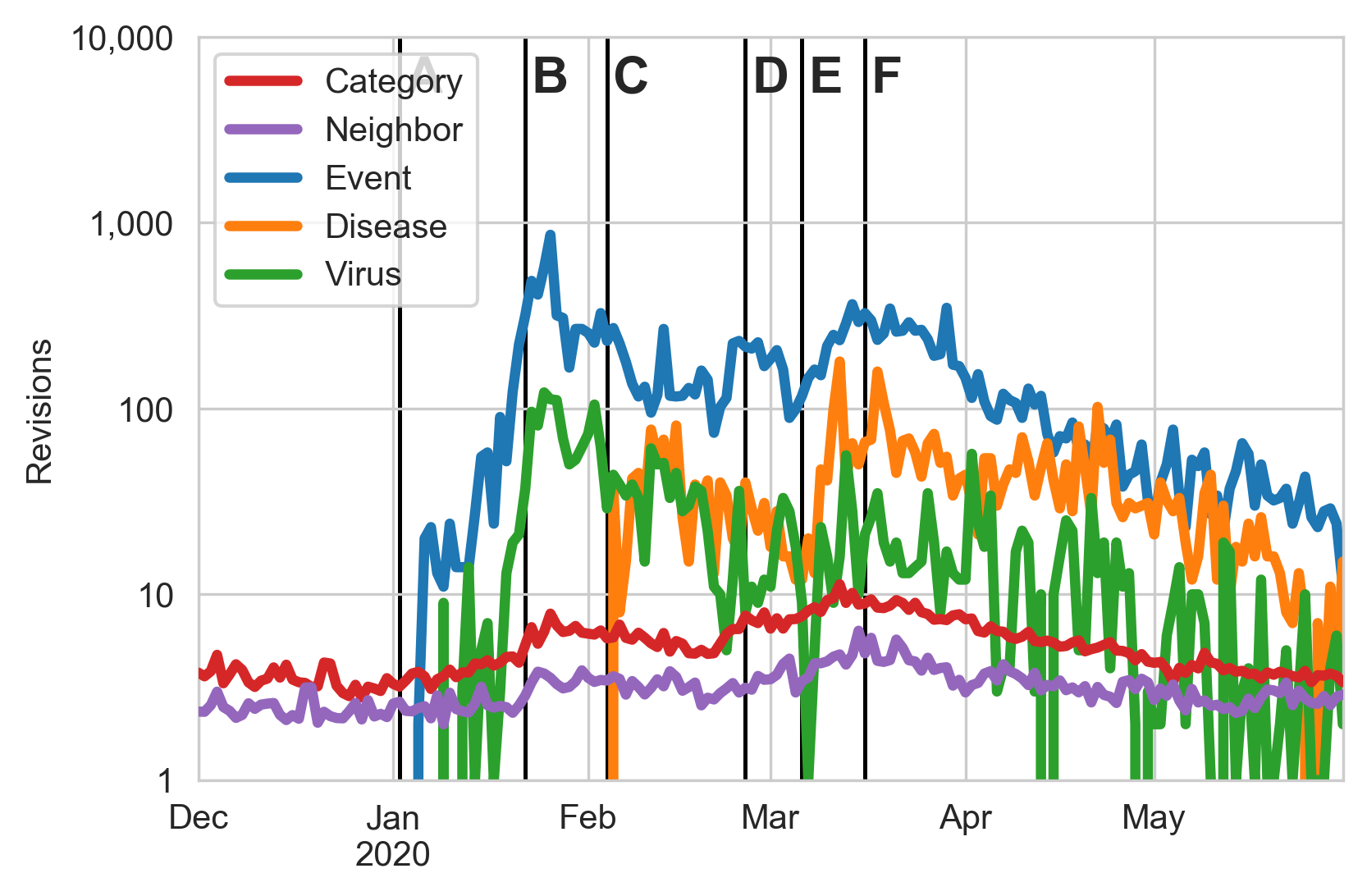}
        \caption{Daily revision counts for the pandemic (blue), disease (orange), and virus (green) articles and mean daily counts for category-related (red) and neighbor-related articles (purple). Events from Table~\ref{tab:timeline} are annotated by letter.
        }
        \label{fig:daily_revisions_seed}
    \end{minipage}\hfill
    \begin{minipage}{.475\textwidth}
        \centering
        \includegraphics[width=\textwidth]{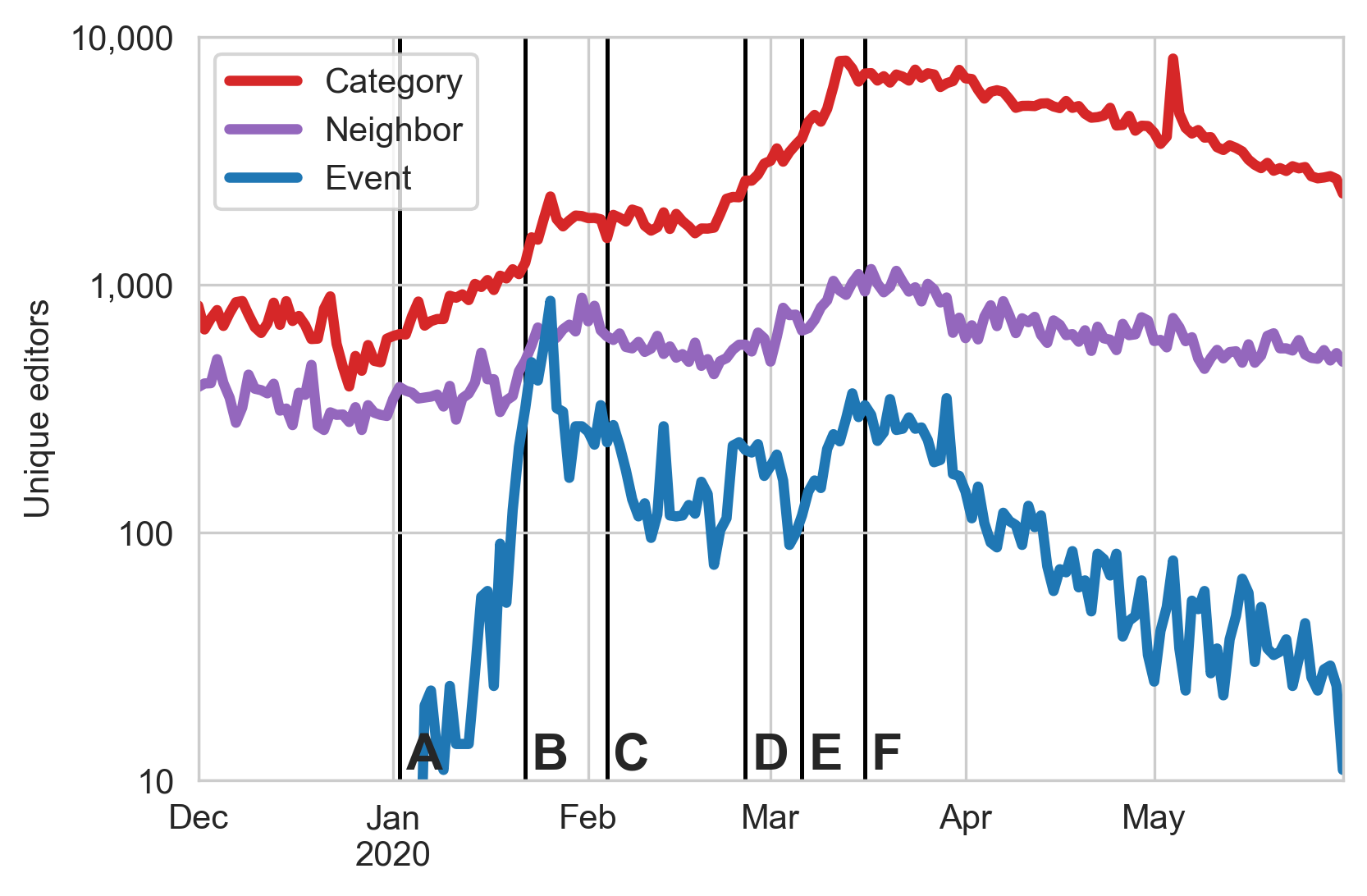}
        \caption{Summed daily revision counts for articles related by category (red) and neighbor (purple) compared to pandemic event article (blue).
        }
        \label{fig:daily_revisions_all}
    \end{minipage}
\end{figure}

How did the number of revisions to articles change over the pandemic? Figure~\ref{fig:daily_revisions_seed} plots the number of daily revisions for the pandemic, disease, and virus articles. The activity on these three seed articles is uneven over time with notable ``bursts'' of activity involving hundreds of revisions in a single day occurring around major events. With so many events unfolding simultaneously around the globe coupled with lags in reporting and article editing, it is difficult to causally identify which specific external events drove the observed bursts of activity. The pandemic event article predates the other two seed articles and has spikes of activity over time windows that correspond with major external events (see Table~\ref{tab:timeline} with events annotated in Figure~\ref{fig:daily_revisions_seed}). The article about the virus was created before the article about the disease, but revision activity on the disease article overtook the virus article. By mid-April, revision activity on all three of these seed articles had fallen below 50 revisions per day, well below their peak activity: 861 revisions to the event article on January 25, 181 revisions to the disease article on March 12, and 123 revisions to the virus article on January 25, indicating the decline of updates and interest on these pages.

The largest spikes of revision activity on the event and virus articles happen late January around the announcements of Chinese quarantines and the first U.S. cases, well before more significant global disruptions in March. A critical but unobserved confounder in these data is that page protections barring ``non-autoconfirmed'' users\footnote{``Auto-confirmed'' accounts are typically created more than 4 days ago and have more than 10 revisions.} from making revisions are placed on the pandemic event article by administrators first intermittently and then indefinitely. Page protection is an important content moderation tool on Wikipedia for preventing disruptions and vandalism, but page protections also prevent new and low-activity accounts from directly participating in the collaborations~\cite{hill_page_2015}. The indefinite page protections introduced after January would have the intended effect of capping the number of accounts qualified to contribute and thus reducing the number of revisions made to articles during the severe escalation in March.

Figure~\ref{fig:daily_revisions_all} plots the number of daily revisions for the pandemic event article and the sum of the daily revisions for the category and neighbor-related articles. Because there are hundreds or thousands of articles within the category and neighbor-related article sets, even low levels of activity distributed across these pages aggregates into more total activity than the primary articles. Category-related pages ($n=3,072$) had over a thousand daily revisions by early January while neighbor-related pages ($n=1,166$) only briefly passed a thousand daily revisions in mid-March. The late-January burst of revision activity on the pandemic event article was matched by bursts of revisions on the category-related articles and a few days later by an uptick in revisions on the neighbor-related articles. From late February through March, activity on the category-related articles more than quadrupled from fewer than 2,000 revisions per day to  8,099 revisions on March 13. 

The burst of revisions to category-related articles in late February through March rather than the primary burst of revisions in January is a {\em much better} match for the likely demand for information than any of the seed articles' protection-censored revision activity suggests. Again, neighbor-related article revision activity surged a few days later with 1,083 revisions on March 15 and peaking at 1,131 revisions on March 21. The lag in revision activity on neighbor-related articles compared to category-related articles could be the result of several mechanisms such as editors prioritizing updates to the most topical content, maxed-out editorial capacity requiring a few days for changes to trickle down to other articles, or consensus to delay updating important neighboring articles like ``United States'' until more reliable sources exist. Finally, while revision activity on the pandemic event article declines after April below pre-peak activity, revision activity on both the category-related and neighbor-related articles remain elevated compared to their pre-peak levels.

\begin{wrapfigure}{R}{.45\textwidth}
    \centering
    \includegraphics[width=.45\textwidth]{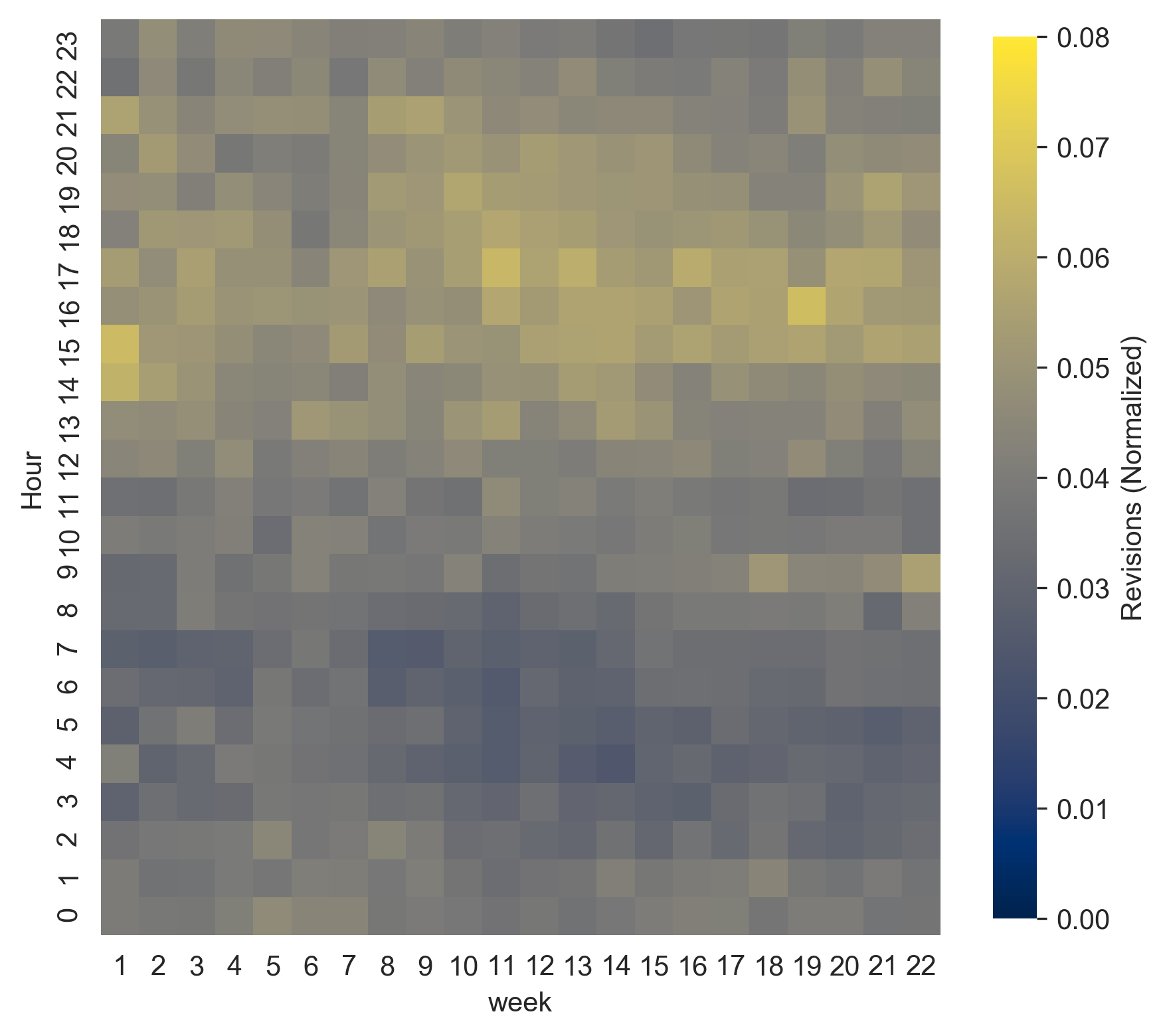}
    \caption{Heat map of the number of unique users making revisions by week (x-axis) and hour of the day (y-axis).
    }
    \label{fig:editing_time_of_day}
\end{wrapfigure}

Figure~\ref{fig:editing_time_of_day} provides another perspective on the revision activity dynamics by plotting how revision activity changed by hour of the day for each week in 2020. Previous research has shown strong diurnal patterns exist in online social behavior~\cite{golder_diurnal_2011,yasseri_circadian_2012}. Despite the global nature of both the event and anglophone Wikipedians, there is nevertheless a strong and consistent diurnal signal in the revision activity of editors: more revisions in the early evening than the middle of the morning. This diurnal pattern persists over time despite profound differences in the size and intensity of the collaboration over time. While ground truth labels on editors' geographic locations are not publicly available, these activity patterns are indicative of a geographic bias in the editors residing within the United States as they edit outside of traditional work hours and sleep hours. This within-day temporal patterning of revision activity has implications for collaboration and coordination if the most active editors are concentrating their attention and activity within narrow windows of time within the day. This points to informal and pluritemporal forms of anticipation, coordination, and delegation exist to manage conflict and to process new information within these collaborations~\cite{cobb_designing_2014,norris_is_2019}.

\subsection{Revision persistence and size}
\begin{figure}[t]
    \begin{minipage}{.475\textwidth}
        \centering
        \includegraphics[width=\textwidth]{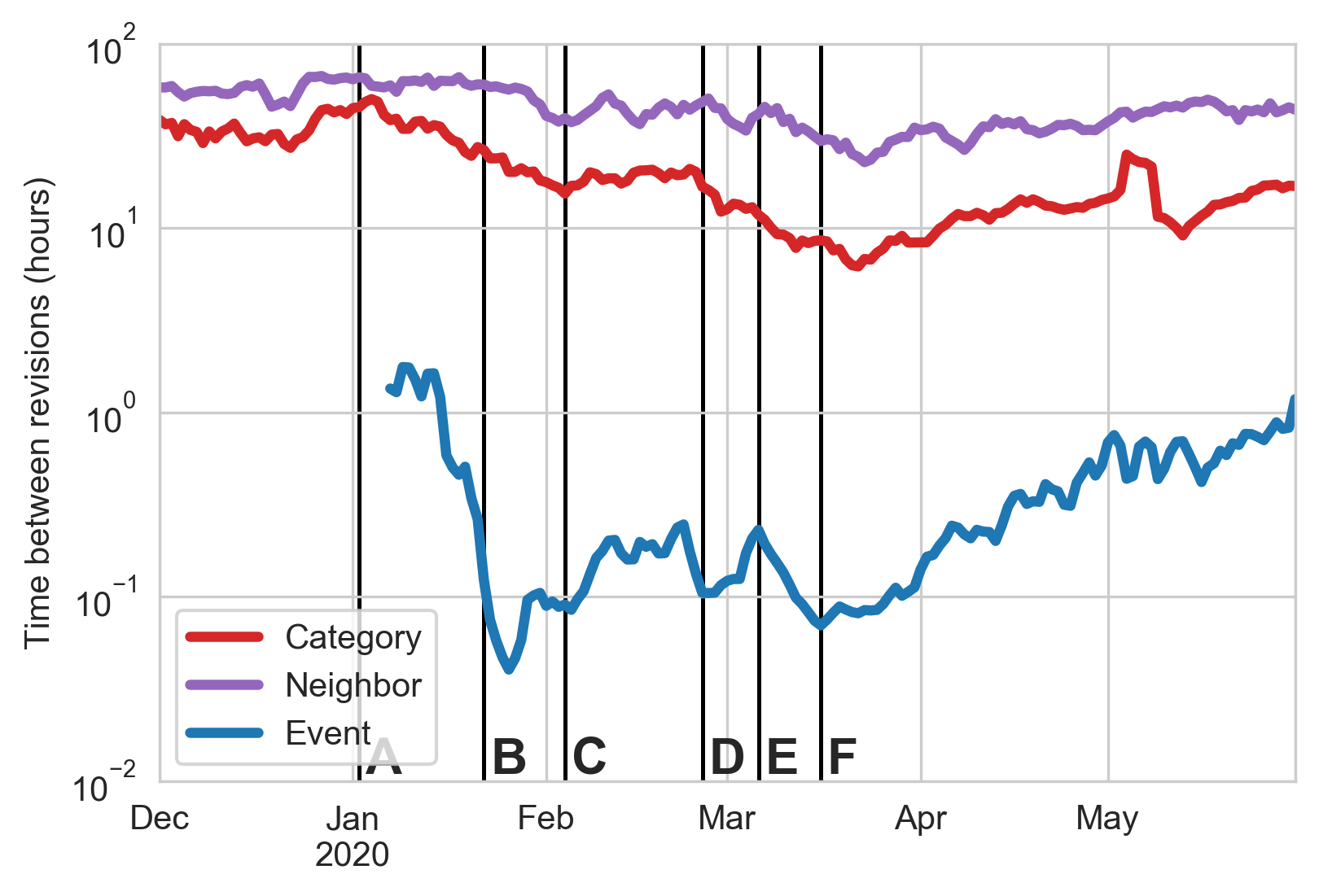}
        \caption{Average time between revisions (in hours) for category-related (red), neighbor-related (purple), and pandemic event (blue) articles.
        }
        \label{fig:daily_edit_lag_all}
    \end{minipage}\hfill
    \begin{minipage}{.475\textwidth}
        \centering
        \includegraphics[width=\textwidth]{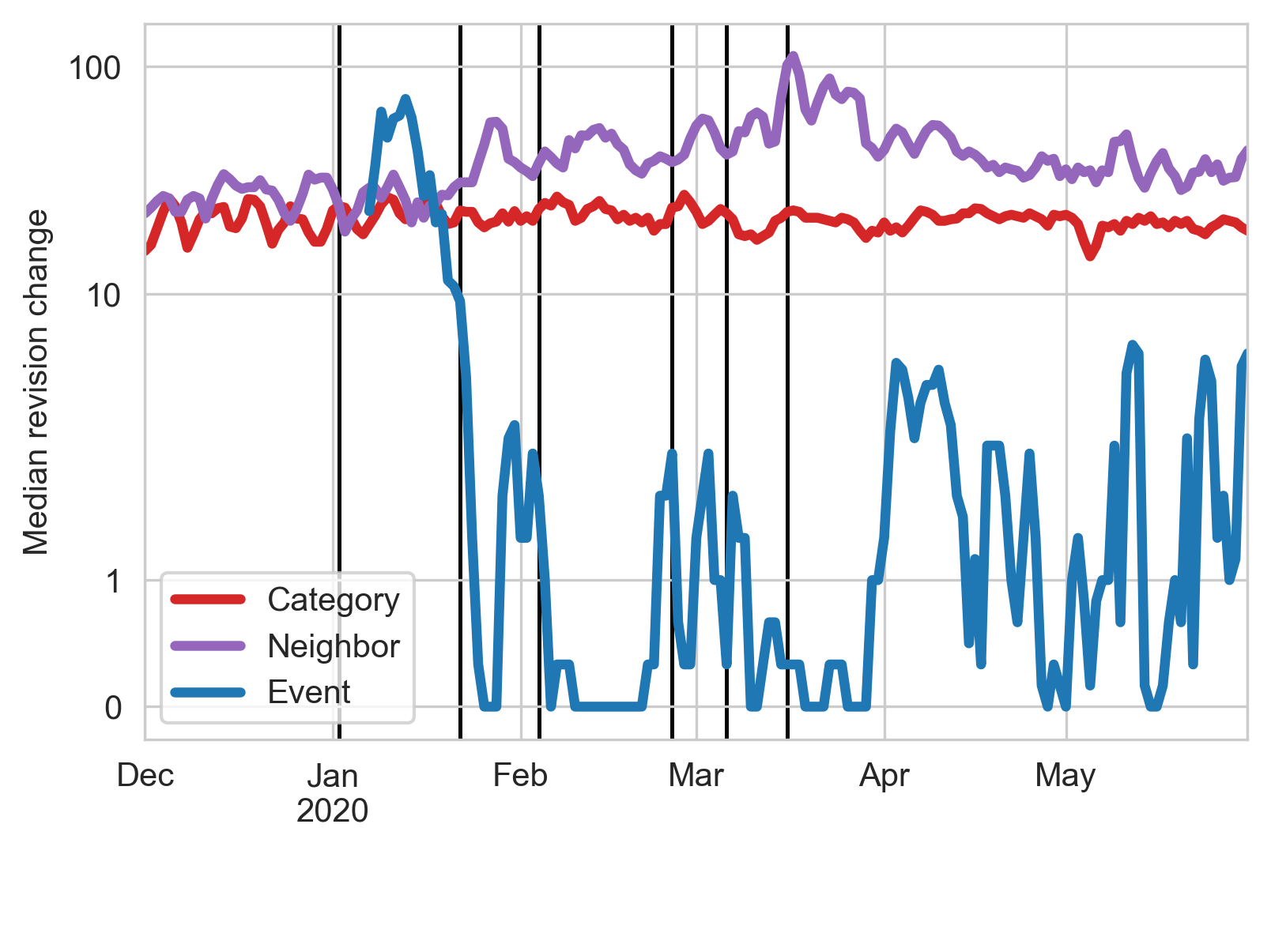}
        \caption{Median size of revisions for category-related (red), neighbor-related (purple), and pandemic event (blue) articles.
        }
        \label{fig:daily_diff_all}
    \end{minipage}
\end{figure}

At least two consequences flow from the significant increases in daily revision activity: the time between revisions must decrease and the size of the revisions should decrease. Figure~\ref{fig:daily_edit_lag_all} plots the average time between revisions (in hours) on the pandemic event article and the category-related and neighbor-related articles. The intense editing activity on the pandemic event article means that any given revision to an article only persists for at most an hour before it is overwritten by another editor's revision. When edit activity peaked in late January, the average revision persistence was only a few minutes. The pace of activity on the pandemic event is much higher than on the average category-related or neighbor-related article (hours versus days of persistence), but both the category-related and neighbor-related articles also see a marked decrease in the persistence in late February and through March before increasing again in April. 

Figure~\ref{fig:daily_diff_all} plots the median size of a revision (in bytes) on the pandemic event article and the two sets of related articles. The MediaWiki software on which Wikipedia runs does not support synchronous editing so edit conflicts can arise when one editor commits a change to the article before another editor finishes theirs. To avoid losing their contributions due to edit conflicts, editors need to change their contribution strategies to reflect a higher-tempo environment. One strategy is to make smaller and more incremental edits, which would manifest as a decrease in the median size of a revision. As expected, median revision size decreases for the pandemic event article from dozens of bytes to 0 bytes during the late-January activity burst. The article briefly rebounds, but returns to 0 median bytes for the rest of February reflecting contributions like incremental copy-editing rather than authoring large new sections. When the pace of events picks up in late February and March, the median edit size becomes sizable again as new content is added and there is less competition to make edits while the page is protected. 
The median revision size increases significantly in early April,
suggesting more reliable information is being produced (hence bigger updates) while edits are slowing down.

The behavior of the category-related and neighbor-related articles are a study of contrasts. Category-related articles see no substantial changes in median revision size over the whole time window. Even as the edit persistence time drops, category-related articles evidently do not become so high-tempo as to require new modes of coordination and editors because editors are consistently able to make larger changes, enabling these articles to grow. Neighbor-related articles' median revision size changes substantially over the time window, from small incremental 1-byte revisions before January to greater than 10-byte changes from late January through March. Their longer edit persistence is one factor enabling larger edits to be made. 
These findings demonstrate the importance of characterizing the revision activity content related to a focal set of articles to better understand how Wikipedia contributors shift their editorial attention over a large information space before, during, and after major events. 

\subsection{Article creation and genealogies}

\begin{figure}[t]
    \begin{minipage}{.475\textwidth}
        \centering
        \includegraphics[width=\textwidth]{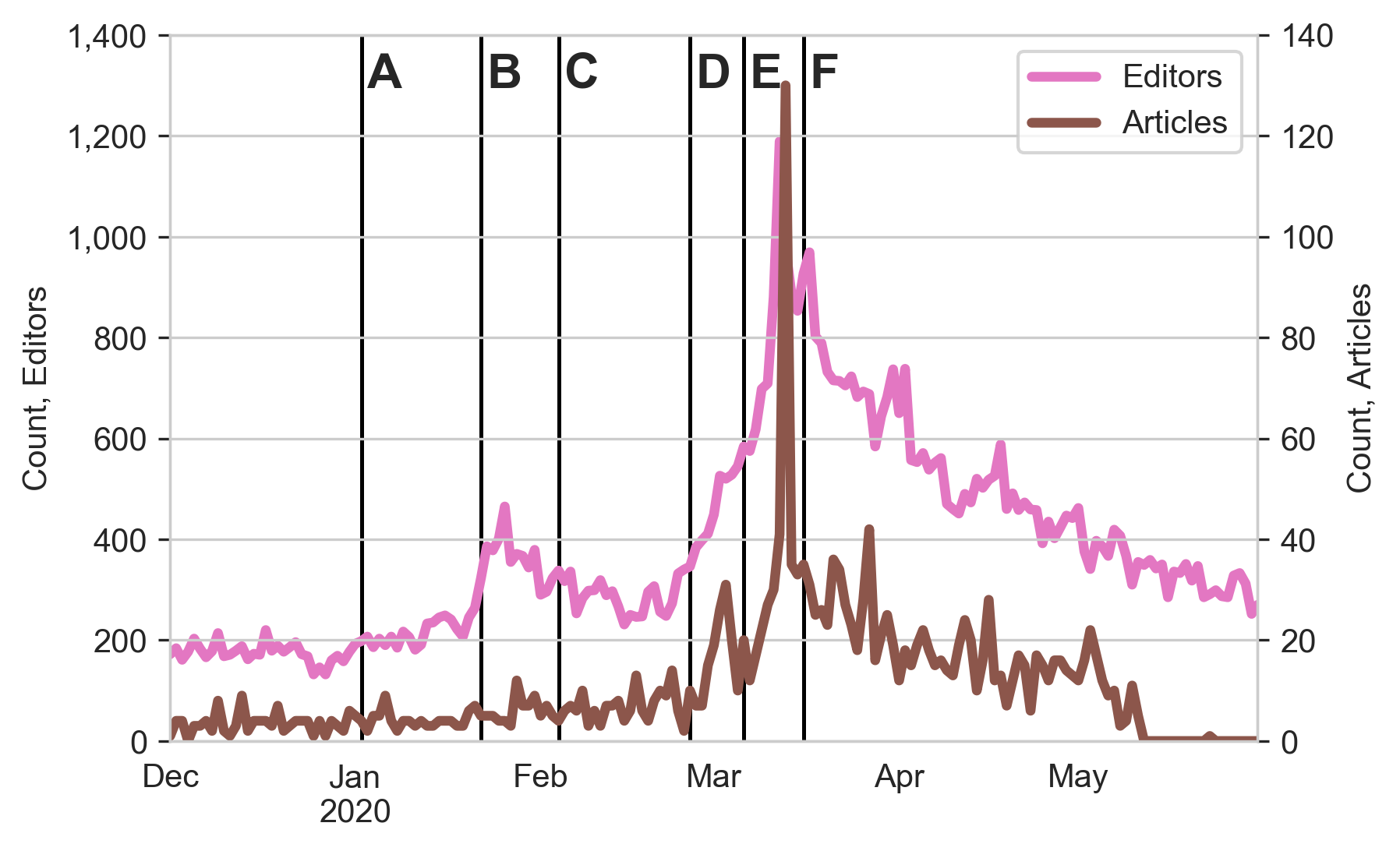}
    \caption{Number of users making their first edit (pink) and articles created (brown) daily.}
    \label{fig:daily_page_creations}
    \end{minipage}\hfill
    \begin{minipage}{.475\textwidth}
        \centering
        \includegraphics[width=\textwidth]{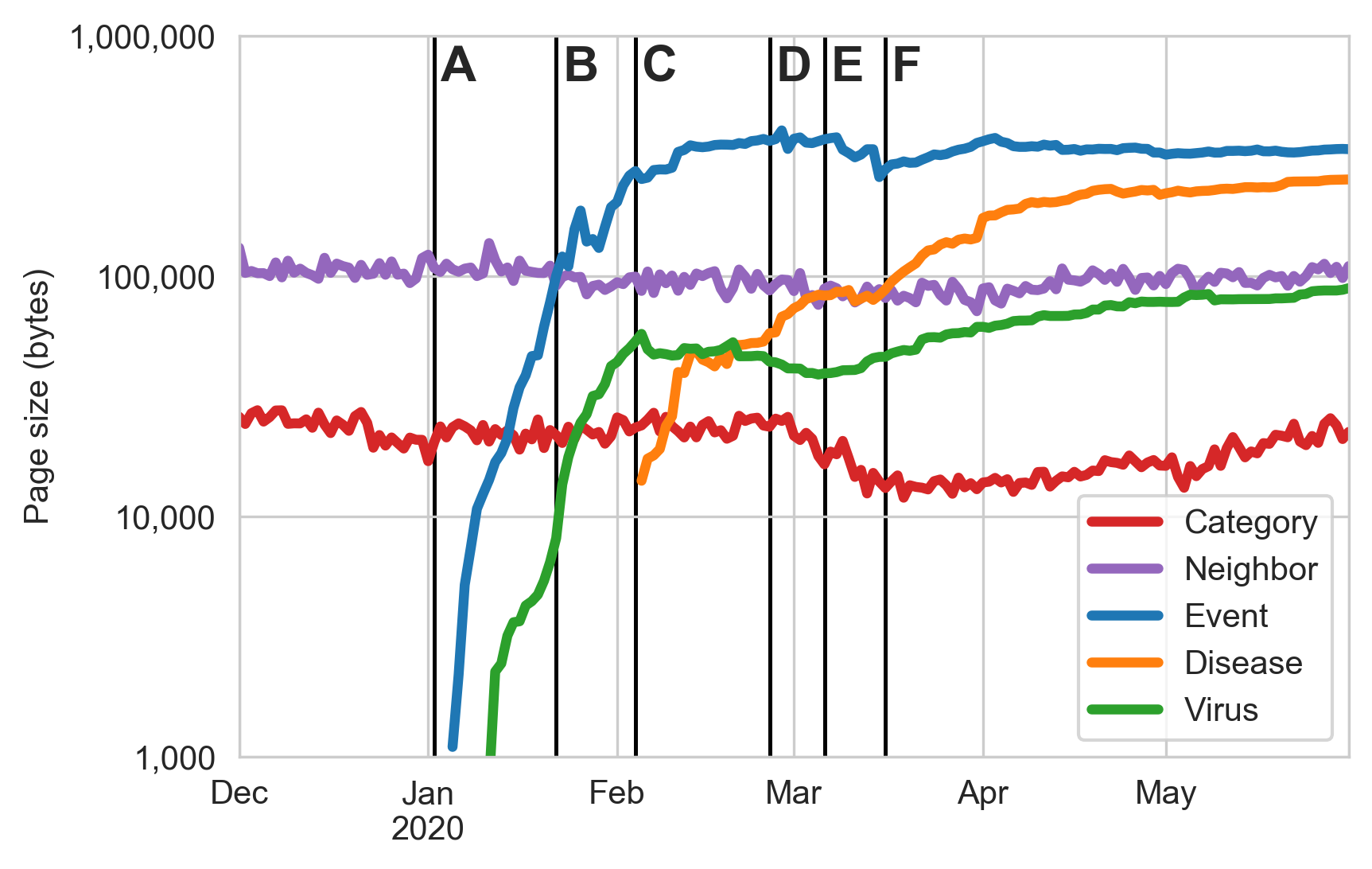}
        \caption{Median daily size (in bytes) for articles related by category (red), neighbor (purple), and individual articles.
        }
        \label{fig:daily_size_all}
    \end{minipage}
\end{figure}

The events surrounding the COVID-19 pandemic require not only updating existing articles but creating new articles. These new articles include both high profile articles like those about the pandemic event, disease, and virus but also includes more mundane new articles like ``Imperial College COVID-19 Response Team'' and ``Graduate Together''. Figure~\ref{fig:daily_page_creations} plots the number of articles created each day within the window in our sample. The article creation activity does not track the category-related article revision activity patterns seen in Figure~\ref{fig:daily_revisions_all} perfectly ($r=0.698$): there is no first peak in late January, a second major peak in mid-March rather than rising in late February, and declining activity through April and May. The prominent peak happens March 12--13 when hundreds of new geographic child articles (``\href{https://en.wikipedia.org/wiki/COVID-19_pandemic_in_Alabama}{COVID-19 pandemic in Alabama}'', ``\href{https://en.wikipedia.org/wiki/COVID-19_pandemic_in_Burundi}{COVID-19 pandemic in Burundi}'', \textit{etc}.) are created. These geographic articles were created by a small handful of editors, initially redirecting to larger geographic entities like ``\href{https://en.wikipedia.org/wiki/COVID-19_pandemic_in_the_United_States}{2020 coronavirus pandemic in the United States}'' or ``\href{https://en.wikipedia.org/wiki/COVID-19_pandemic_in_Africa}{2020 coronavirus pandemic in Africa}'' before the redirect was removed and the article populated with standalone content days or weeks later. 

Figure~\ref{fig:daily_page_creations} also captures the number of editors who made their first revision to any of the articles in the combined sample. 
The timing of editors' first engagements with these collaborations tracks overall category-related article revision activity very well ($r=0.940$). 
However, the top ten most active editors within the sample began their revision activity to articles in this sample before 2020; they were already making revisions to articles that were later linked to by the pandemic event article or migrated into the category-related articles. The topics of these most active editors' first revisions to articles in the sample are not obvious predictors of editors' intense activity during the pandemic response: ``\href{https://en.wikipedia.org/wiki/2019\%E2\%80\%9320_EHF_Champions_League}{2019--20 EHF Champions League}'', ``\href{https://en.wikipedia.org/wiki/Sepsis}{Sepsis}'', ``\href{https://en.wikipedia.org/wiki/2020_Caribbean_Club_Championship}{2020 Caribbean Club Championship}'', ``\href{https://en.wikipedia.org/wiki/Singapore_Changi_Airport}{Singapore Changi Airport}'', and ``\href{https://en.wikipedia.org/wiki/Supergirl_(TV_series)}{Supergirl (TV series)}''. Previous analyses of Wikipedia's high-tempo online collaborations have found similar patterns of editors migrating their expertise from prosaic but related articles to current events articles~\cite{keegan_hot_2011,keegan_staying_2012,keegan_posthumous_2015,twyman_black_2017}. What motivates these editors to migrate to current events collaborations remains an open question.

\begin{figure}[t]
    \centering
    \includegraphics[width=.9\textwidth]{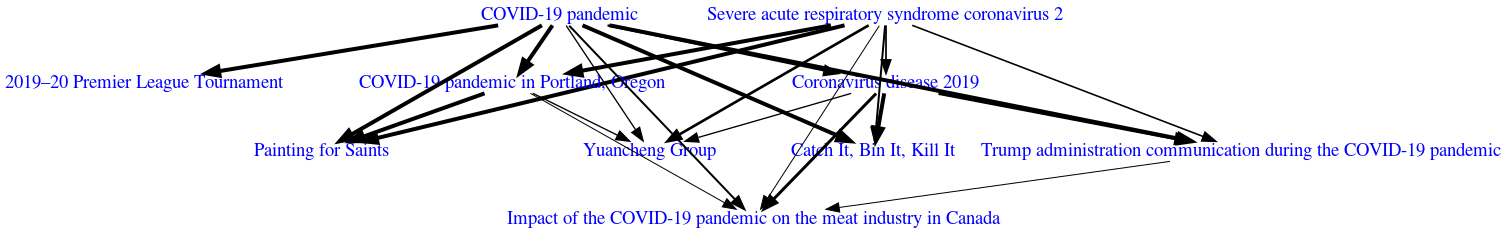}
    \caption{Visualization of the genealogical graph of the three seed articles and their top three children.
    }
    \label{fig:genealogy}
\end{figure}

Another way of exploring the processes of article creation is to employ a genealogical framework. New articles are not created in isolation by novices, but are the product of dedicated and/or disaffected users who move their contributions from a previous article to a new article: article creations are also editor migrations. We employ a method for building genealogical graphs within online communities by connecting the communities (in this case, individual Wikipedia articles) together if the earliest editors of new articles were also editors previously contributing to other articles~\cite{tan_tracing_2018}. Note these are distinct from collaboration networks, because a directed relationship only exists if the ``parent'' article and its activity predates the ``child'' article. These ``family trees'' of online communities could play a role in the success by recruiting committed members or ensuring access to important resources~\cite{cunha_are_2019}. Figure~\ref{fig:genealogy} visualizes the backbone of the strongest genealogical relationships around the three seed articles. Some of these genealogical relationships are unsurprising (the disease article is a child of the virus article) but others reveal hidden flows of editors moving from general to more specific or unexpectedly related topics.

\subsection{Article size and content similarity}
Figure~\ref{fig:daily_size_all} plots the changes in the size of articles (in bytes) over the time window. The median size for the neighbor-related articles are large and relatively stable: these high-profile articles (see Table~\ref{tab:articles}) tend to be old, long, and active with only minor changes needed to be updated for the COVID-19 pandemic. The median size for the category-related articles are much smaller but decrease significantly from 22,652$\pm$2,708 bytes on average to 14,908$\pm$2,450 bytes ($F=385.9, p<0.001$) after the mid-March article creation burst (Figure~\ref{fig:daily_page_creations}). This is an artifact of there being hundreds of smaller ``stub'' articles, like the geographic child articles, bringing the median article size down. The median size of these category-related articles remains depressed throughout April and May, reflecting low levels of revision activity on these articles (Figure~\ref{fig:daily_revisions_seed}).  

The category-related and neighbor-related article sizes also provide context for understanding the scale and intensity of the collaborations on the seed articles. The pandemic event article was created on January 5 at 1,098 bytes and grew extremely rapidly in size, passing the category-related articles' median size on January 15 and neighbor-related articles' median size on January 22. The disease article was created much later on February 5, but had a very large initial commit (14,063 bytes), and passed the category-related articles' median size on February 10 and neighbor-related articles' median size on March 4. The virus article was created on January 9, started small (776 bytes), and grew more slowly, passing the category-related articles' median size on January 25, although its size 
has not yet passed the median neighbor-related article size. The rapid growth in the length of the pandemic event and disease articles is remarkable: they overtake the detailed neighbor-related articles within weeks. The pandemic event article has an apparent ceiling on its size: the article's size was 323,885$\pm$49,807 bytes in February and 334,932$\pm$23,200 bytes after March, a non-significant difference ($F=2.34,p=0.129$). The event article's content obviously did not stagnate (Figures~\ref{fig:daily_revisions_seed}) but rather saw content being replaced  (Figure~\ref{fig:daily_diff_all}) as sections were moved to child articles and new citations, media, and templates were added. The disease article has shown slower but consistent growth since its creation only the beginning of a ceiling while the virus article plateaued in February but has also started to grow in length since mid-March.

\begin{figure}[t]
    \centering
    \includegraphics[width=\textwidth]{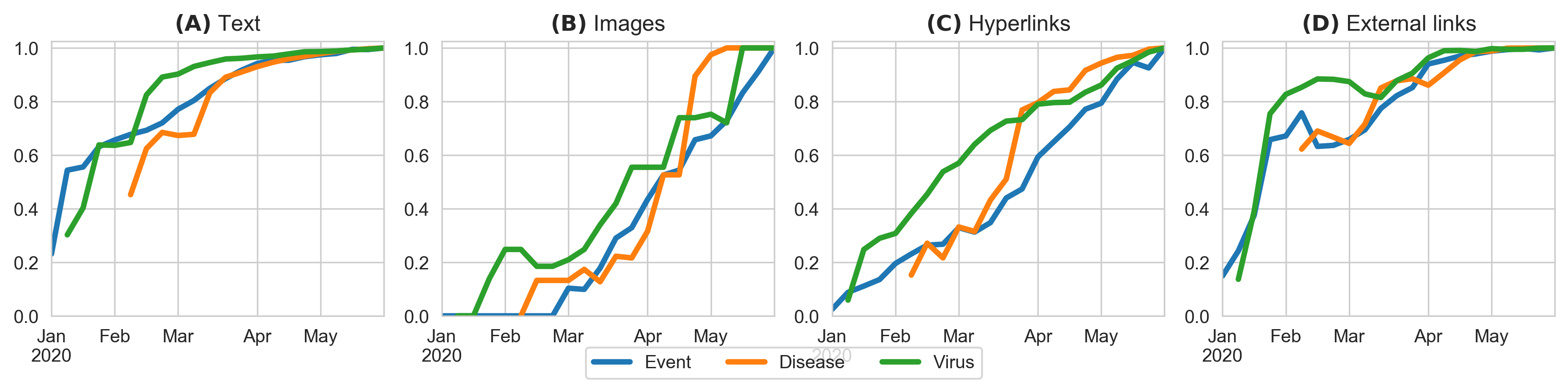}
    \caption{Changes in the similarity of the text (A), images (B), hyperlinks (C), and external links (D) of the pandemic (blue), disease (orange), and virus (green) articles compared to the most recent week.
    }
    \label{fig:weekly_seed_content_changes}
\end{figure}

The sizes of the seed articles do not capture the dynamics of how their content has changed over time. Figure~\ref{fig:weekly_seed_content_changes} plots changes over time in the similarity of four content features on the three seed articles compared to the most recent version of each article in our data.

\begin{description}[leftmargin=1em]
\item[Textual similarity] (Figure~\ref{fig:weekly_seed_content_changes}A) is measured as the cosine similarity between a previous and the most recent revisions' unigrams in the body of the article (excluding text in image captions and templates) after pre-processing for common English stopwords and punctuation. The pandemic event article shows a pattern of consistently increasing similarity, reflecting on-going but incremental changes in the text rather than sudden shifts. The disease and virus articles have distinctive ``steps'' in their textual similarity over time, reflecting major additions or edits of content. Text similarity of all articles is relatively high (>0.9) before April, implying that the structure of these articles stabilized quickly and the revision activity is only making incremental changes to the text.
\item[Image similarity] (Figure~\ref{fig:weekly_seed_content_changes}B) is measured as the cosine similarity between a previous and the most recent revisions' counts of large images (>100 pixel width to exclude icons). There is a much lower level of similarity in early versions of the articles indicating that few of the images currently present in the article were included then. The consistent change in image similarity week-over-week captures on-going turnover rather than stabilization in the articles' images.
\item[Hyperlink similarity] (Figure~\ref{fig:weekly_seed_content_changes}C) is measured as the cosine similarity between a previous and the most recent revisions counts of hyperlinks. Like images, there is an initially low level of similarity and consistent week-over-week growth in similarity reflecting on-going turnover rather than stabilization in the articles' hyperlinks.
\item[External link similarity] (Figure~\ref{fig:weekly_seed_content_changes}D) is measured as the cosine similarity of the normalized counts of top-level domains present in external links between a previous revision and the latest revision. Examples include links to the \texttt{bbc.com}, \texttt{who.int}, \texttt{cdc.gov}, and \texttt{nytimes.com}. Like text, early versions of the article have relatively high similarity to the current version (>0.6) and changes over the subsequent weeks are incremental. The content dynamics here show a plateau in similarity scores for February into March, reflecting an stabilizing consensus.
\end{description}

\noindent The content on these seed articles show different dynamics and provide additional context explaining the revision activity (Figure~\ref{fig:daily_revisions_seed}), revision size (Figure~\ref{fig:daily_diff_all}), and article size (Figure~\ref{fig:daily_size_all}). The text and length of these seed articles may not have changed significantly since March, but there is still a significant amount of on-going turnover in the choice of images, hyperlinks, and external links. The changes within a type of content are similar across the three articles, implying the dynamics of expanding and sustaining the content of new articles about current events may follow consistent pathways. The rapid stability in the textual similarity of the article suggests that early editors' efforts to frame the article may be extremely durable over a time-span of weeks while high-tempo collaborating demands other editors make more incremental changes that manifest in the consistent and significant changes in the similarity of images and hyperlinks. Page protection again is a critical unobserved confounder~\cite{hill_page_2015} limiting the kinds of editors who can contribute content. Previous researchers have characterized this collaborative inertia as territoriality~\cite{thom_mine_2009,thom_territoriality_2010}, but opportunities for larger-scale editing and reappraisal may open on these articles, as it does for other Wikipedia articles that are subject to high-tempo collaborations~\cite{keegan_posthumous_2015,twyman_black_2017}, after a few months.

\subsection{Pageviews}
\begin{figure}[t]
    \begin{minipage}{.475\textwidth}
        \centering
        \includegraphics[width=\textwidth]{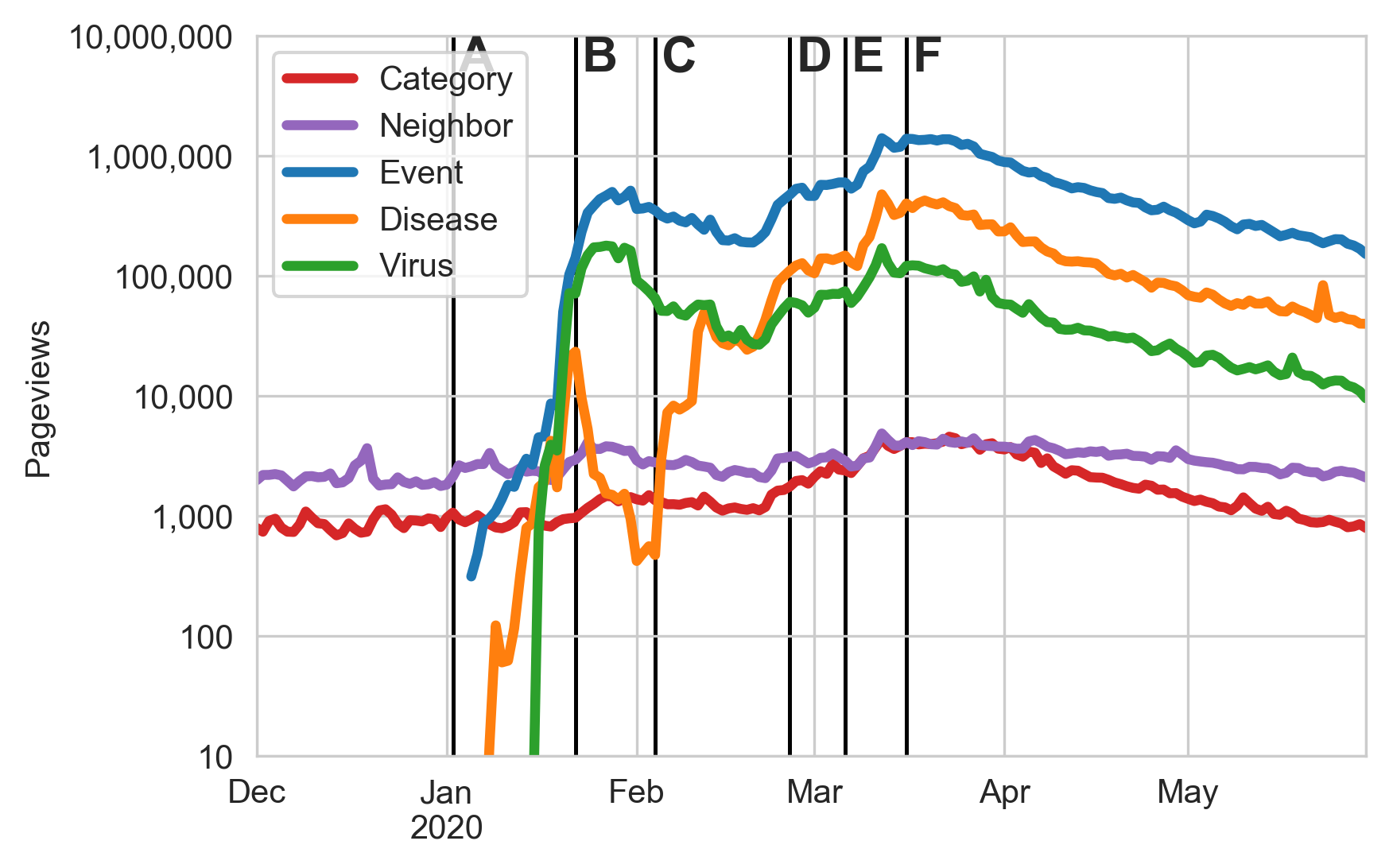}
        \caption{Daily pageviews for the pandemic (blue), disease (orange), and virus (green) articles with median values for category-related (red) and neighbor-related (purple) articles.
        }
        \label{fig:daily_pageviews_seed}
    \end{minipage}\hfill
    \begin{minipage}{.475\textwidth}
        \centering
        \includegraphics[width=\textwidth]{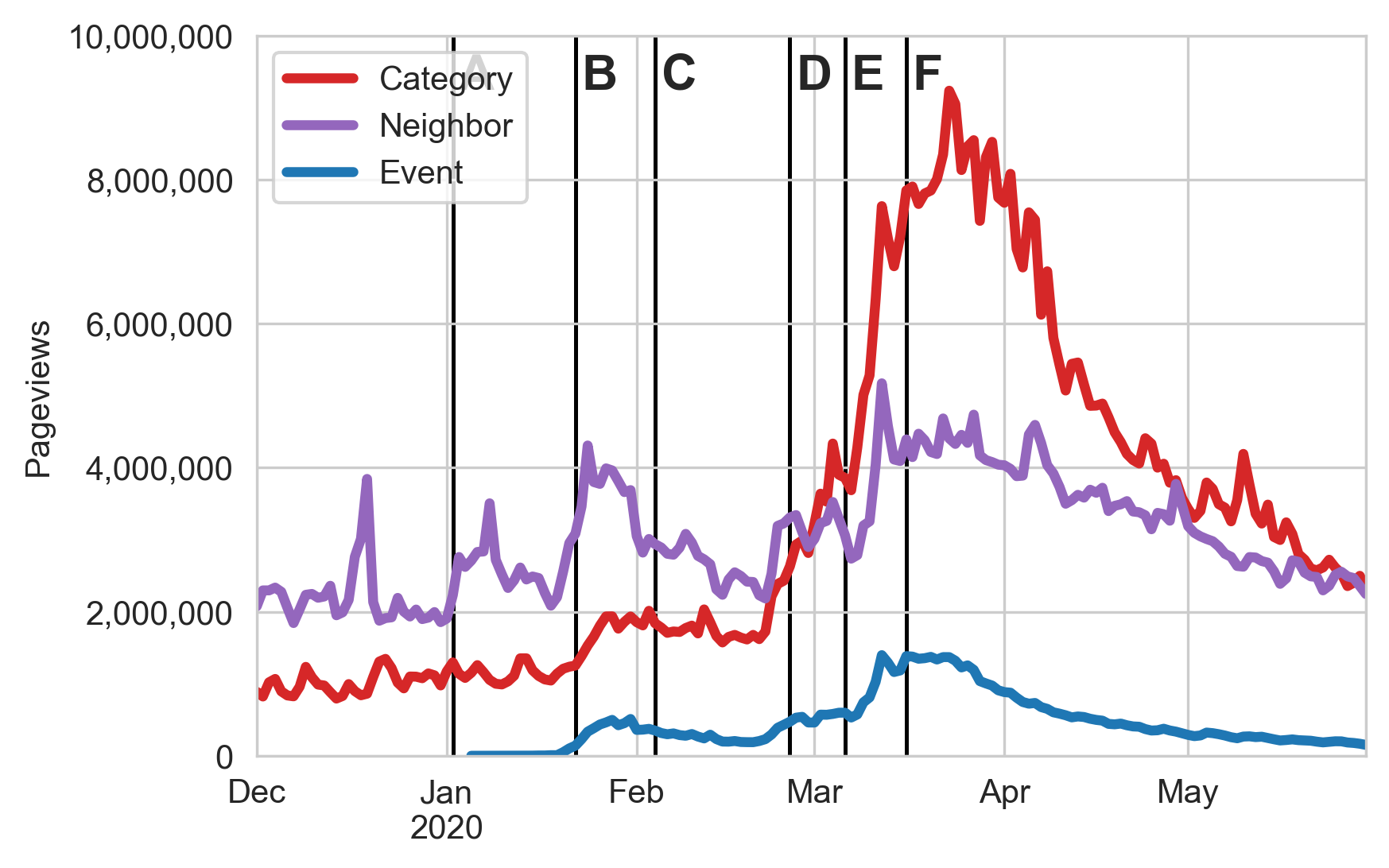}
        \caption{Daily pageviews for articles related by category (red), neighbor (purple), and pandemic (blue).
        }
        \label{fig:daily_pageviews_all}
    \end{minipage}
\end{figure}

The previous article-level analyses have focused on the dynamics around the \textit{production} of content, but Wikipedia's pageview statistics provide a (coarse) measure of the \textit{consumption} of this content over time. As we noted in the description of our approach, the Wikimedia Foundation does not disaggregate pageview data by geography, device, \textit{etc}. for privacy reasons. Figure~\ref{fig:daily_pageviews_seed} plots the daily pageviews for the three seed articles and the median pageviews for articles in the set of category-related and neighbor-related articles. The event and virus pages show a characteristic ``double spike'' of attention centered on the late-January quarantines in China and the mid-March quarantines in Europe and the United States followed by a consistent decline in attention. Because the disease article was not created until February 5, there was no first peak on this article. But pageviews to non-existent pages are still logged and reported and there actually is a small but perceptible burst of a few hundred pageviews in late January for the yet-to-be-created disease article that redirected at the time to the pandemic event article. If extremely actively-viewed articles like ``Donald Trump'' or ``United States'' receive approximately 50,000 pageviews daily, the relative level of attention to these three pandemic-related seed articles during this time window is profound. The last time the pandemic event article had fewer than 50,000 pageviews was January 19 and the disease article on February 22. The maximum pageviews on the pandemic event article passed 1,000,000 for several days in March but still remains in excess of 100,000 daily.

The pageviews for the neighbor-related articles ($\bar{pv}_{pre}=2501\pm573, \bar{pv}_{post}=3511\pm636, F=113, p<0.001$) and category-related articles ($\bar{pv}_{pre}=1090\pm303, \bar{pv}_{post}=2670\pm1024, F=197, p<0.001$) increase modestly but significantly after March 1. Median pageview activity on category-related articles is always less than on neighbor-related articles, again reflecting the greater relative prestige of articles like ``New York City'' than ``COVID-19 pandemic in New York City''. But the correlated increases of attention to category-related ($r=0.965$) and neighbor-related ($r=0.869$) articles is an example of an ``attention spillover'' as people navigate to or from the seed articles from the category-related and neighbor-related articles driving up their pageview numbers as well~\cite{dimitrov_what_2017,kummer2018attention,lamprecht_how_2017,west_mining_2015,zhu2018content}. Similar spillovers in editorial attention were seen in Figures~\ref{fig:daily_revisions_seed} and~\ref{fig:daily_pageviews_all} when revision activity on related articles increased significantly as editors updated these articles with new information. While attention spillovers---particularly to the neighbor-related articles---could also be driven by other exogenous events unrelated to the pandemic (\href{https://en.wikipedia.org/wiki/2019\%E2\%80\%932020_Persian_Gulf_crisis}{2019–2020 Persian Gulf crisis}, \href{https://en.wikipedia.org/wiki/2020_Democratic_Party_presidential_primaries}{2020 Democratic Party presidential primaries}, \textit{etc}.), observing a significant increase across articles representing a diverse set of topics (public health, transportation, politicians, \textit{etc}.) gives us greater confidence that the correlated bursts of attention on related articles are pandemic-related information seeking.

Figure~\ref{fig:daily_pageviews_all} plots the daily pageviews for the pandemic event seed article as a baseline and the sum of pageviews for all articles in the category-related and neighbor-related sets. While the median pageviews of category-related articles never overtook the neighbor-related articles (Figure~\ref{fig:daily_revisions_seed}), there is an unmistakable surge of demand for information about category-related topics in March that overtakes all the neighbor-related articles. The pandemic event article makes up only a small fraction of this surge and while the remaining attention may have been split across more than 3,000 category-related articles, this captures the enormous demand for information as the severity of the pandemic and its social effects accelerated. Measuring event-driven attention spillovers, the neighbor-related articles see a much clearer increase for the January events than the category-related articles as well as the characteristic second peak. The absence of this first peak for the category-related articles can likely be attributed to the lack of child articles about impacts, locations, and response created in February and beyond (Figure~\ref{fig:daily_page_creations}). 

\section{Editor-level activity}\label{sec:user}

Using the editors who contribute to articles as the primary unit of analysis, we examine how the size of these seed articles collaborations grew over time, the patterns of user re-engagement, edit session length, and changes from editors' previous contribution behavior.

\subsection{Re-engagement}
\begin{figure}
    \centering
    \begin{minipage}{.475\textwidth}
        \centering
        \includegraphics[width=\textwidth]{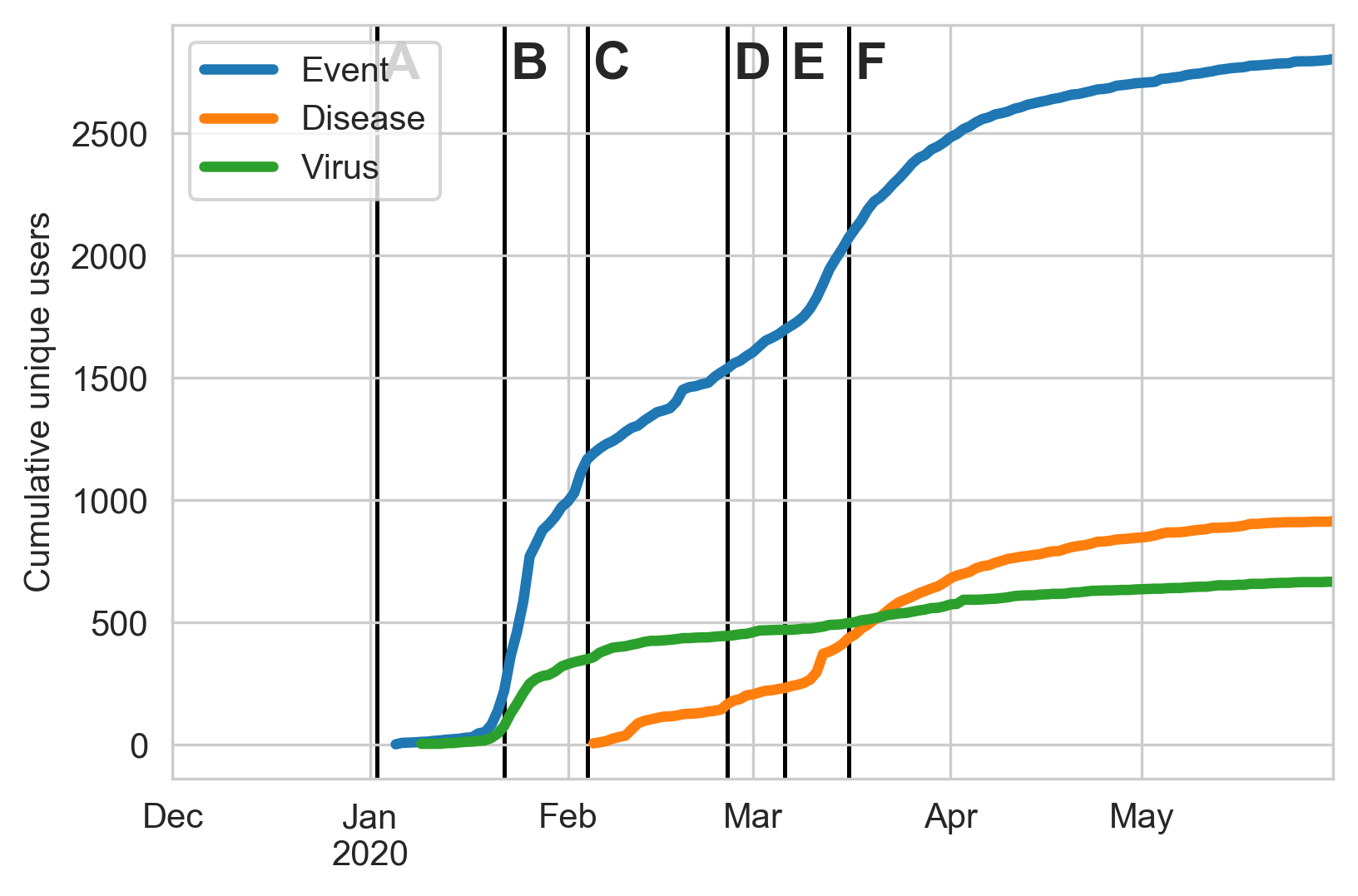}
        \caption{Cumulative number of unique contributors on pandemic event (blue), disease (orange), and virus (green) articles.
        }
        \label{fig:daily_cuml_users_seed}
    \end{minipage} \hfill
    \begin{minipage}{.475\textwidth}
    \includegraphics[width=\textwidth]{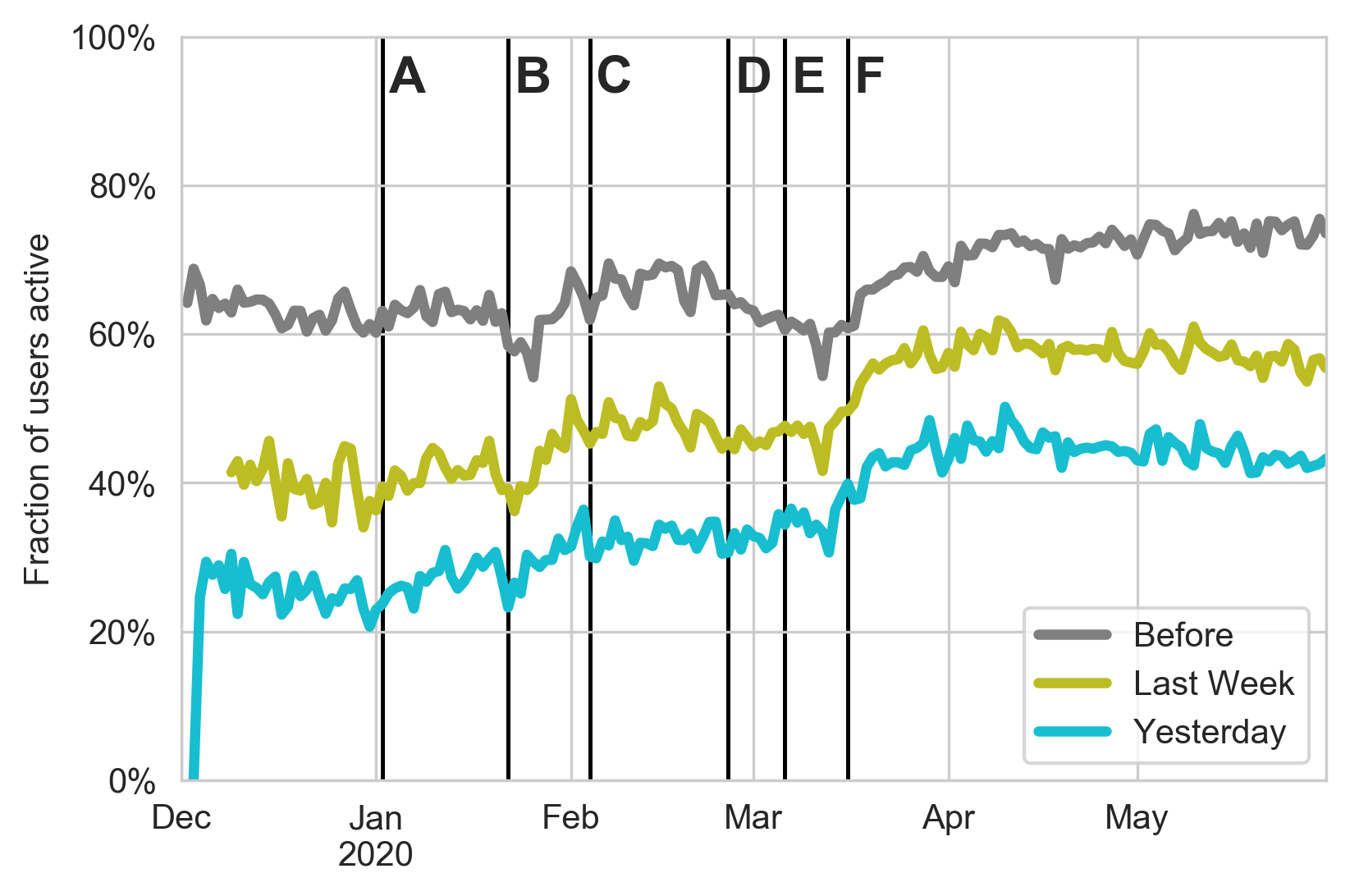}
    \caption{Fraction of users active who also made a revision to any in-sample article (category and neighbors) at any time before (grey), in the last week (olive), and yesterday (cyan).
    }
    \label{fig:daily_user_reengagement}
    \end{minipage}
\end{figure}

The number of editors active daily on an article closely tracks the total number of daily revisions (Figure~\ref{fig:daily_revisions_seed}). Figure~\ref{fig:daily_cuml_users_seed} plots the cumulative number of unique editors who contributed to each of the three seed articles by day. Over the December 2018 through May 2020 sampling window, articles in our sample had a median of 25 and an average of 57.8 unique editors. The seed articles accumulated significantly more editors in a few months than these other articles had in years: the pandemic event article 2,749 unique editors, the disease article 881 editors, and the virus article 646 editors. Changes in the number of editors on these articles correspond with major turning points in the pandemic. This rapid growth in the number of contributors to these articles could be a source of tension as editors bring different values and expertise, but it is also the case that most editors' engagement on these articles is extremely limited, typically making only one revision to one article (see Figure~\ref{fig:cuml_collab_g_distributions}A and C and discussion in next section). 

\begin{figure}[t]
    \centering
    \includegraphics[width=\textwidth]{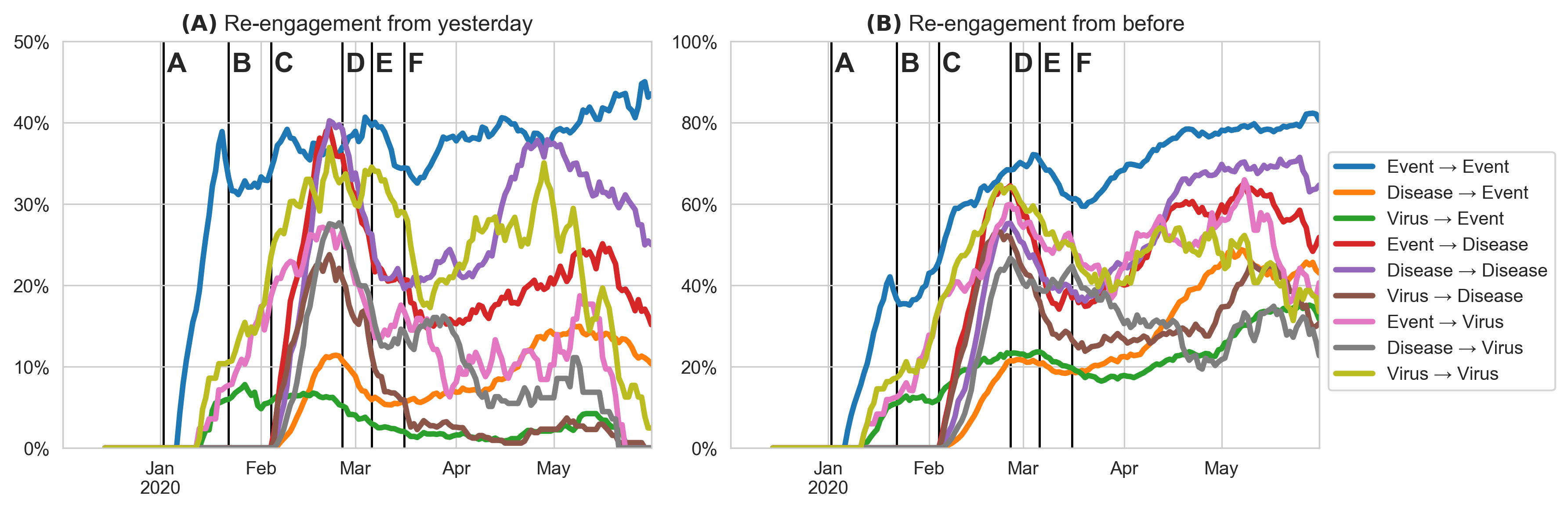}
    \caption{Percentage of users active on the event, disease, or virus article who were active yesterday (left) or any time before (right) on the event, disease, or virus articles.
    }
    \label{fig:daily_user_reengagement_seeds}
\end{figure}

How many of the editors active on an article one day were previously active? Figure~\ref{fig:daily_user_reengagement} plots the fraction of editors who made a revision to any category-related or neighbor-related article between December 2019 and May 2020. On any given day, at least 20\% of the editors also made a contribution yesterday and at least 60\% of the editors made a contribution in the previous year. The re-engagement of editors in making contributions is not static and there is a significant increase after March 1 in editors' re-engagement ($re$) from yesterday ($\bar{re}_{pre}=28.4\pm4.7\%, \bar{re}_{post}=42.4\pm4.8\%, F=346.4, p<0.001$), in the last week ($\bar{re}_{pre}=43.3\pm4.2\%, \bar{re}_{post}=55.3\pm5.0\%, F=259.1, p<0.001$), and in the past year ($\bar{re}_{pre}=64.0\pm2.9\%, \bar{re}_{post}=68.7\pm5.1\%, F=54.9, p<0.001$). Even as the size and intensity of the collaboration surges, a plurality of the contributing editors have previously been active on these pages. These high levels of re-engagement indicate a strong motivation among editors to monitor and participate in the collaboration rather than treating them as ``drive by'' contributions. Even as overall revision (Figure~\ref{fig:daily_revisions_seed}) and pageview (Figure~\ref{fig:daily_pageviews_seed}) activity has declined through April, these re-engagement statistics remain at their significantly elevated levels.

\begin{wrapfigure}{r}{.35\textwidth}
    \centering
    \includegraphics[width=.35\textwidth]{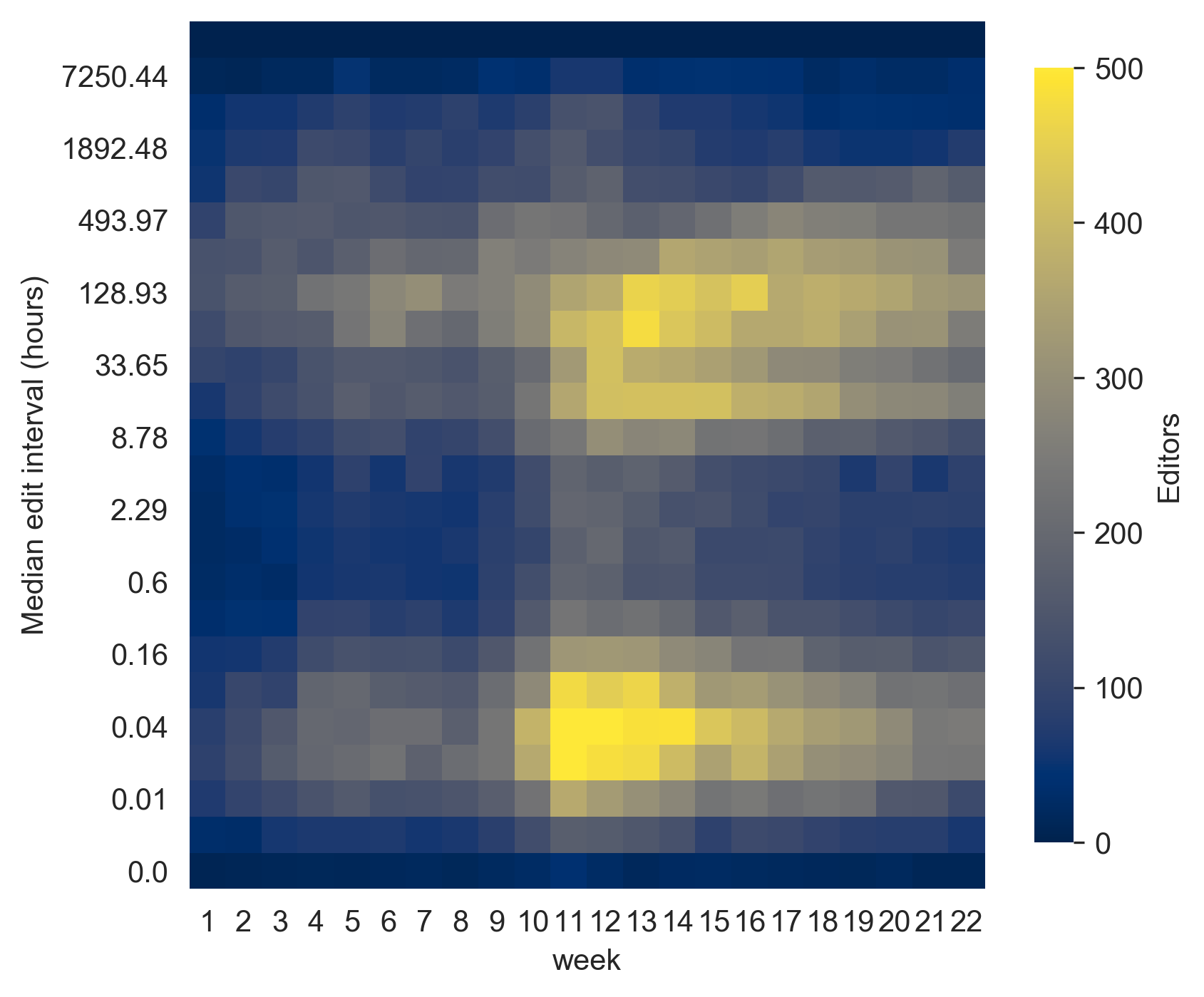}
    \caption{Changes in the distribution of users' daily median revision latencies.
    }
    \label{fig:editing_intensity}
\end{wrapfigure}

The re-engagement statistics reported in Figure~\ref{fig:daily_user_reengagement} are across the whole sample of articles and does not capture whether editors are ``staying in their lane'' contributing on their usual articles or how they are moving between articles. Figure~\ref{fig:daily_user_reengagement} visualizes the re-engagement dynamics of editors moving within and between the three seed articles as measured day-over-day and from anytime since article creation. For example, 30--40\% of editors on the pandemic event article on any given day after mid-January were also active on the previous day (blue line, Figure~\ref{fig:daily_user_reengagement}A). In mid-February, close to 60\% of the editors active on the disease article were active on the event article the day, falling to 40\% in mid-March, and rising again to more than 60\% in early May (red line, Figure~\ref{fig:daily_user_reengagement}A). The February 5 creation of the disease article is an occasion for significant collaborative migrations of editors from other articles to contribute to the new article. The intensity of cross-article re-engagement declines in early March with the surge of activity related to external events bringing in newcomers. However, these newcomers leave or are converted into committed editors and the re-engagement numbers recover rising through April and into May. These cross-article re-engagement dynamics are a critical mechanism for coordination and standardizing content across articles while also illustrating a remarkably strong commitment among editors for stewarding their contributions and participating in these collaborations for an event lasting months.

\subsection{Edit sessions}
Previous work exploring edit sessions on Wikipedia identified a characteristic double-peak in contribution patterns: one reflecting behavior occurring in relatively rapid sequences with only minutes between successive revisions and another reflecting daily patterns with approximately a day between revisions~\cite{geiger_sessions_2013}. Given the significant, sudden, and uneven shifts in revision activity, attention, content, and re-engagement around pandemic articles, it is reasonable to hypothesize that stable patterns of edit sessions could not hold. Similar to the idea of edit lags within an article's revision history in Figure~\ref{fig:daily_edit_lag_all}, we can also analyze the time between revisions in a user's contribution history within this sample of category-related and neighbor-related articles. 

Figure~\ref{fig:editing_intensity} plots a two-dimensional histogram of users' median revision latencies changing over time. The y-axis are logarithmic bins of time between observed revisions, ranging from seconds at the bottom to months at the top. The x-axis are weekly bins of activity since January 1. The values in each cell are the number of users in that week who had a median revision latency in that bucket's range. We find a similar characteristic double-peak in revision latencies that is consistent over time during even this high-tempo collaboration. There is one population of users who typically make multiple revisions only minutes apart and second population of users who typically make revisions once every few days. The number of editors in each of these populations increases in week 11 corresponding to the early March surge of activity, but the double-peak remains consistent over time.


\subsection{Cohort analysis}
\begin{figure}[t]
    \centering
    \includegraphics[width=\textwidth]{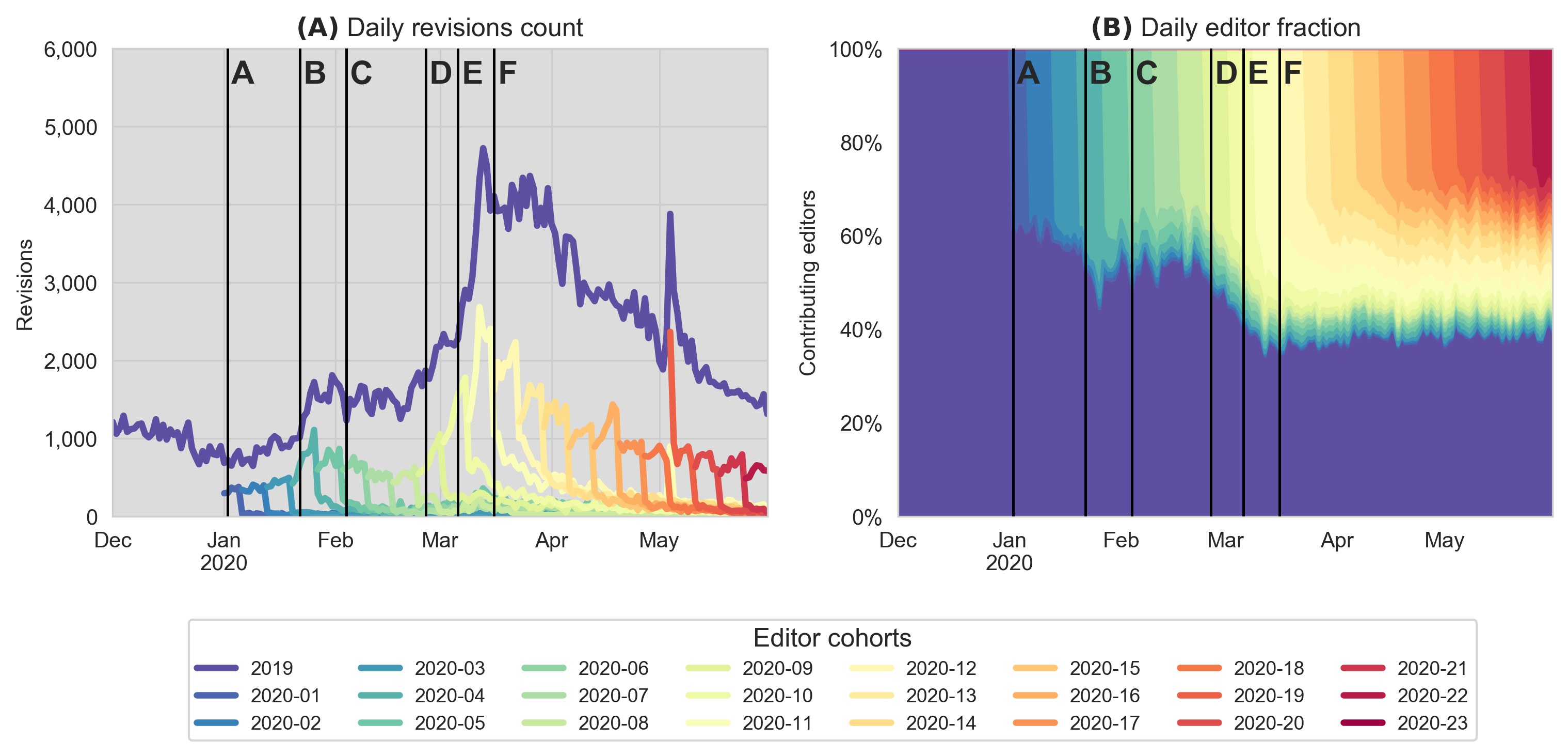}
    \caption{The number of daily revisions to all articles from different editing cohorts (left) and the percentage of users active on all articles from different editing cohorts (right). Editing cohorts are weekly in 2020 and all users first active before 2020 are labelled 2019.}
    \label{fig:cohort_daily_activity}
\end{figure}

Simpson's paradox arises when heterogeneous data is aggregated producing different associations and trends than exist in any of the underlying subgroups and is an endemic problem in the social and computational sciences~\cite{blyth_simpsons_1972,lerman_computational_2018}. One strategy for surfacing the existence of this paradox are ``time-aware analyses'' by bucketing users into cohorts to explore whether differences exist in the behavior of these subgroups~\cite{barbosa_averaging_2016,alipourfard_using_2018}. We group editors into cohorts based on their first contribution to articles in our sample: a ``2019'' cohort for all users who made contributions before 1 January 2020 ($n=60,258$) and weekly cohorts for users making their first in-sample contributions in 2020 ($n=58,296, \bar{n}_{i}=2,915$). Figure~\ref{fig:cohort_daily_activity}A revisits the findings from Figure~\ref{fig:daily_revisions_all} by plotting the number of daily revisions from editors in each of these cohorts. The revision activity from the ``2019'' editors are the most active, partially because they are the largest cohort, but also because it includes highly-engaged editors. In contrast to this 2019 cohort, the number of revisions from later cohorts shows a consistent pattern of temporary engagement: these are the newcomers who make some contributions to in-sample articles but most are not found a week later. This replicates previous findings about the significant churn in editors of politicians' biographies and campaign articles~\cite{keegan_presidential_2019}.

Figure~\ref{fig:cohort_daily_activity}B plots the daily percentage of editors making revisions drawn from each of these cohorts. Even the editors who began contributing during the mid-March surge had only temporary engagement with the article. Obviously revisions made before 2020 can only come from the 2019 cohort, but this 2019 cohort is responsible for more than 40\% of the active users in 2020. Again, the 2020 cohorts' temporary engagement manifests as a single or few days of contributions that become negligible within a week or two. These results reproduce similar findings about high levels of turnover in editor cohorts on Wikipedia articles~\cite{keegan_presidential_2019}, but they cast a new light on the findings from Figures~\ref{fig:daily_user_reengagement} and~\ref{fig:daily_user_reengagement_seeds} that implied a strong tendency for users to remain engaged day-over-day and week-over-week. The 2019 cohort of editors includes a subset of editors whose sustained and consistent revision activity over months is responsible for the majority of revisions and editors during the surges of activity (\textit{e.g.}, mid-March) as well as sustaining maintenance  (\textit{e.g.}, late April). 
This finding highlights the critical role of established contributors.

\subsection{Editor backgrounds}
\begin{figure}
    \centering
    \begin{minipage}{.475\textwidth}
        \centering
        \includegraphics[width=\textwidth]{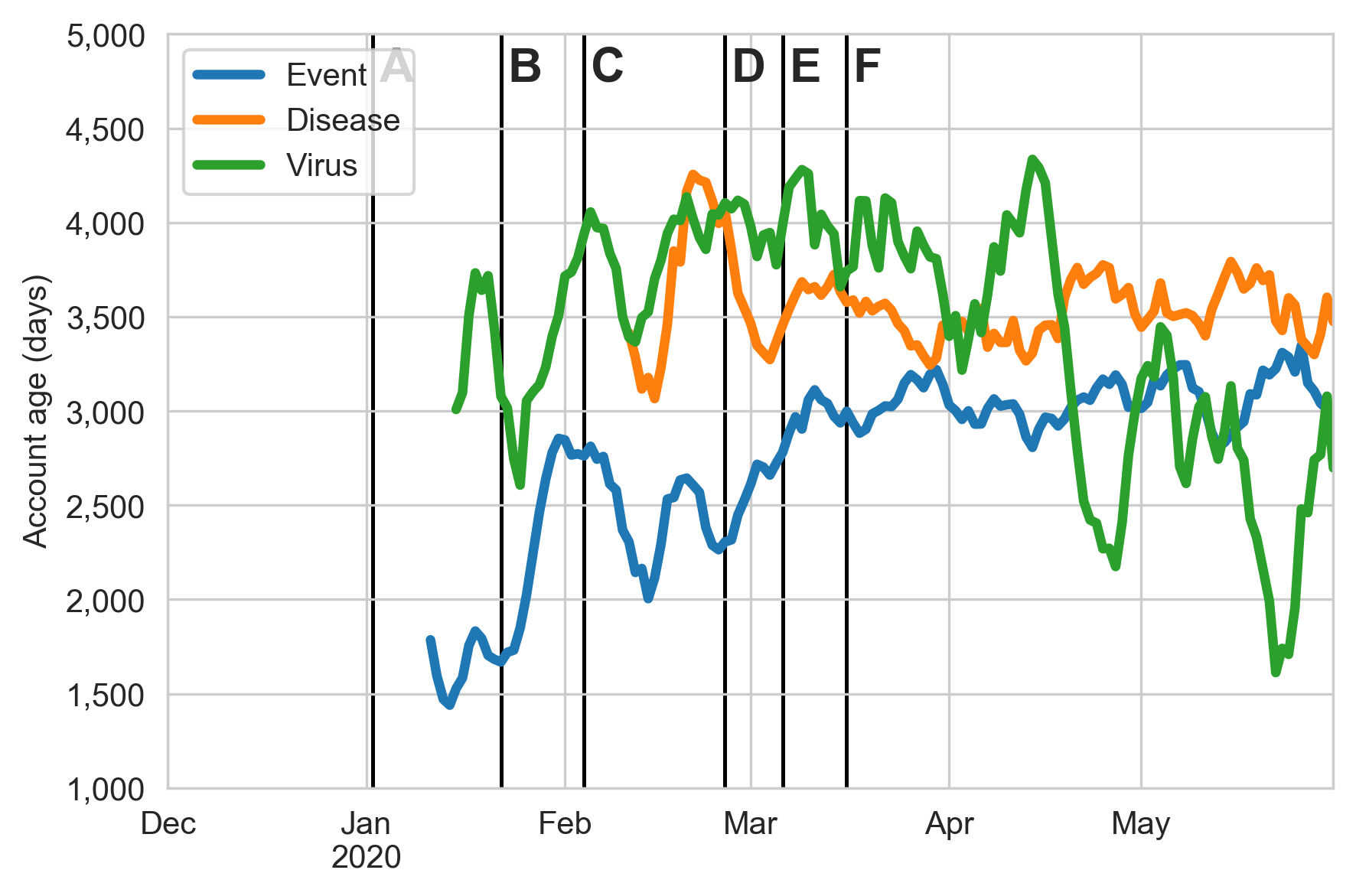}
        \caption{Revision count-weighted average age (in days) of registered editor accounts on the pandemic event (blue), disease (orange), and virus (green) articles.
        }
        \label{fig:daily_account_age_seeds}
    \end{minipage} \hfill
    \begin{minipage}{.475\textwidth}
    \includegraphics[width=\textwidth]{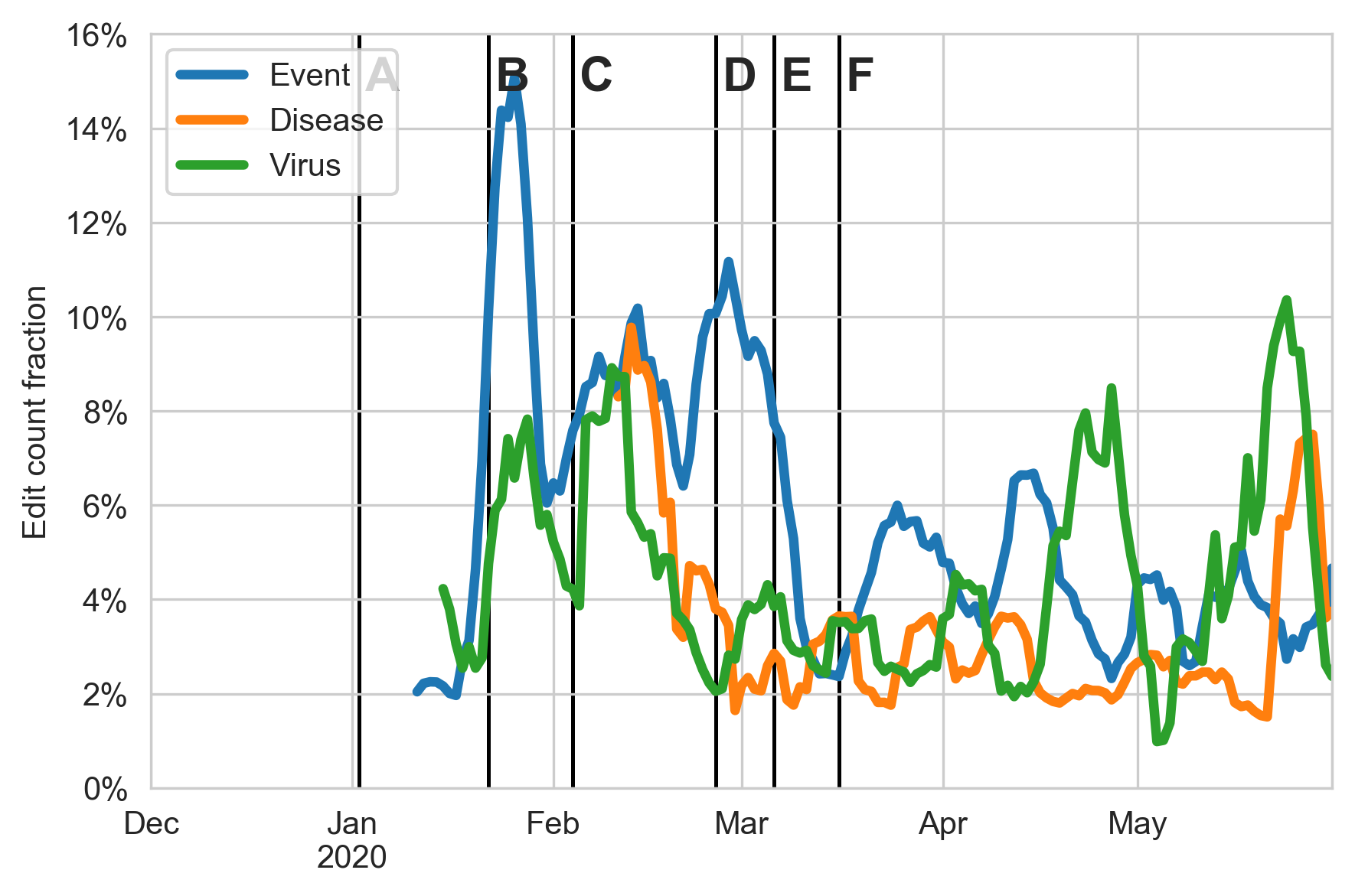}
    \caption{Revision count-weighted average edit count of registered editor accounts on the pandemic event (blue), disease (orange), and virus (green) articles.
    }
    \label{fig:daily_edit_count_fraction_seeds}
    \end{minipage}
\end{figure}

What are the backgrounds of editors on the seed articles? There are many ways to explore this question and we do it through two analyses. For each registered editor who contributed to these three seed articles, their summary statistics including total global edit count and account creation date were retrieved from the ``userinfo'' API endpoint. These user-level statistics were joined with the revision histories to compute  revision-count weighted aggregations: if an editor with an account age of 1,000 days and a 1,000 editcount made 10 revisions to a seed article on a day, their account age and edit count would be counted 10 times in that day's average statistic. Because edit counts can be strongly skewed by editors using automated tools, we employ a normalization dividing the total revisions an editor made to any of the seed articles by their total edit count. This ``edit count fraction'' can be thought of as the total focus an editor had on these seed articles over their whole editing career: editors who contributed only to coronavirus-related topics and nothing else would have an edit count fraction of 1.0, editors made 50 revisions to articles about penguins than then made 50 revisions to coronavirus-related topics would have an edit count fraction of 0.50.

Figure~\ref{fig:daily_account_age_seeds} plots revision-count weighted average age of registered editors on the three seed articles. The early editors on the pandemic event article before February had relatively young accounts (less than 2,000 days) but the average account age increases to approximately 2,750 days with the first editing surge in late January, rises again to approximately 3,000 days with the second editing surge in March, and remains high thereafter. The other seed event articles have much higher average editor account ages over the time window but some similar increases in account age correlated with external events. The increases in account age reflect an influx of more established editors to these articles. 

Figure~\ref{fig:daily_edit_count_fraction_seeds} plots the revision-count weighted average editor count fraction on the three seed articles. 
Before approximately March 1, the editors making contributions to these articles had high edit count fractions indicating a high degree of attention and effort focused on these coronavirus collaborations. The values for all three seed article decrease with the March surge as established editors whose efforts were previously focused elsewhere enter the collaboration, a finding consistent with the increase in account ages seen in Figure~\ref{fig:daily_account_age_seeds} as well as the persistence of the 2019 cohort of editors seen in Figure~\ref{fig:cohort_daily_activity}. Whether this hand-off of contribution between two distinct groups of editors, from the younger founders to an older guard, was accompanied by conflict or relief is a question for future research.

\subsection{Behavioral change}
\begin{table}[t]
    \footnotesize
    \centering
    \begin{tabular}{lcccc}
        \toprule
        \textbf{Feature} & \textbf{Before} & \textbf{After} & \textbf{ANOVA} & \textbf{Kruskal-Wallis} \\ \midrule
        \textit{Dates active} & $26.9\pm29.5$ & $29.3\pm30.9$ & $6.76, p=0.009$ & $9.4, p=.002$ \\
        \textit{Latency} & $7.46\pm29.5$ & $3.79\pm16.0$ & $29.2, p<0.001$ & $0.0, p=0.855$ \\
        \textit{Namespaces} & $3.89\pm3.1$ & $4.13\pm3.3$ & $6.95, p=0.008$ & $5.0, p=0.026$ \\
        \textit{Revisions} & $366\pm1903$ & $386\pm1117$ & $0.21, p=0.646$ & $3.1,p=0.077$ \\ \midrule
        \textit{Category revisions} & $8.46\pm51.4$ & $43.5\pm178.8$ & $90.5, p<0.001$ & $1727, p<0.001$ \\
        \textit{Neighbor revisions} & $9.80\pm45.9$ & $5.19\pm21.0$ & $21.36, p<0.001$ & $59.0, p<0.001$ \\ \textit{Neither revisions} & $1147\pm3390$ & $337\pm1050$ & $132.8, p<0.001$ & $319, p<0.001$\\ 
        \bottomrule
         &  \\
    \end{tabular}
    \caption{Summary of editors' changes in revision behavior after their first revision to any of the seed articles. Mean and standard deviation are reported for each feature for activity before and after the first revision and statistical tests and p-values are reported for differences.}
    \label{tab:behavioral_change}
\end{table} 
\begin{figure}
    \centering
    \includegraphics[width=\textwidth]{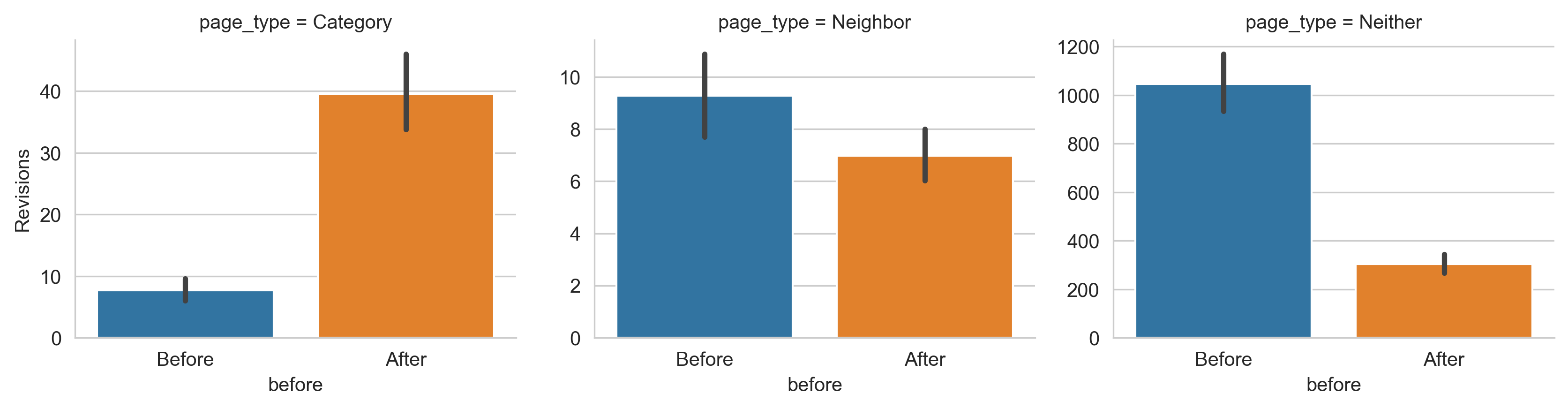}
    \caption{Visualization of changes in editors' revision activity across article types from Table~\ref{tab:behavioral_change}.}
    \label{fig:seed_users_pagetype_prepost}
\end{figure}

Did participating in these breaking news collaborations change editors' contribution patterns? Using the set of editors who contributed to the three seed articles, their contribution histories from December 2018 to the present were retrieved from the Wikipedia API. Using these editors' first contributions to any of the three seed articles as a pseudo-treatment, we match the amount of time in each editors' contribution history (across all articles and namespaces, not just our COVID-19 sample) from their first revision to their most recent revision with an equivalence amount of time immediately before the first contribution: if an editor was active for 30 days after their first revision to one of the seed articles, we look at their entire contribution history 30 days before and after that first contribution. We perform simple statistical analyses using both ANOVA (``are means the same?'') and Kruskall-Wallis (``are medians the same?'') to test whether contributing to one of these seed articles changed users' overall contribution behavior within matched time windows around their first contribution to a seed article.

Table~\ref{tab:behavioral_change} reports the results of these analyses. After their first revision to a seed article, editors make contributions on significantly more days, significantly less time elapses between their revisions, they contribute to more namespaces (articles, talk pages, administrative processes, \textit{etc.}), but do not make significantly more revisions. This last result is surprising because participating in an engrossing breaking news collaborations could stimulate editors to contribute more frequently, if only because they have to make more and smaller edits (Figure~\ref{fig:daily_edit_lag_all}) and editor re-engagement increases through our time window (Figure~\ref{fig:daily_user_reengagement}). A follow-up analysis of editors' changes in revision activity to category-related, neighbor-related, and unrelated articles revealed significant differences in revision activity with important implications. Editors' revision activity on category-related articles increased significantly---five-fold---after their first revision to a seed article, but revision activity on both neighbor-related articles and unrelated articles \textit{decreased} significantly: editors shift their focus from neighbor-related and unrelated articles to category-related articles but do not increase their activity after contributing to a seed article. 

One worrying implication of this profound decrease in activity on unrelated articles is that breaking news collaborations may be draining editors' attention away from other parts of Wikipedia. Whether other editors step in to pick up the slack on these forsaken articles, these editors return back their previous editing routines in the medium term, or these editors burn out and take a break after participating in a breaking news collaboration remain open research questions. Additionally, the non-zero before values for category-related and neighbor-related articles imply editors both (1) contribute more to neighbor-related articles \textit{before} contributing to seed articles and (2) contribute more to category-related articles \textit{after} contributing to seed articles. This suggests an aggregate flow of editors recruiting themselves from both unrelated and neighbor-related articles to join the seed article collaborations and then moving on to other category-related articles afterwards. 

\section{Network dynamics}\label{sec:network}

Networks are aggregations of relational data and network dynamics capture how the nodes and edges within a network change over time. These types of networks are not exhaustive of all the relevant networks---discussion, citation, and categories are examples of other relationships that warrant future analysis---but we focus on two to understand the movement of editors and changes in article content. First, we explore the \textit{collaboration} networks of editors making revisions to articles and how these structures changed as editors shifted their contributions across an expanding set of relevant articles. Second, we explore the \textit{hyperlink} networks of articles linking to other articles and how these structures changed as new information came to light. 
We define each of these networks in more details in the sub-sections below.

\subsection{Collaboration}

\begin{figure}[t]
    \centering
    \includegraphics[width=\textwidth]{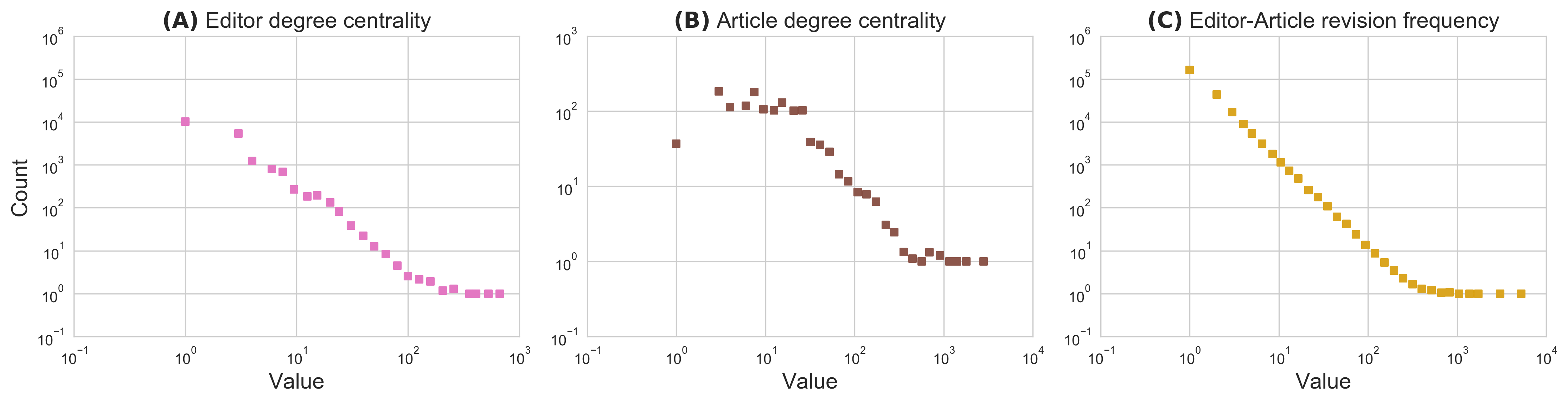}
    \caption{Cumulative collaboration network activity (logarithmically-binned) distributions.
    }
    \label{fig:cuml_collab_g_distributions}
\end{figure}

\begin{figure}[t]
    \begin{minipage}{.475\textwidth}
        \centering
        \includegraphics[width=\textwidth]{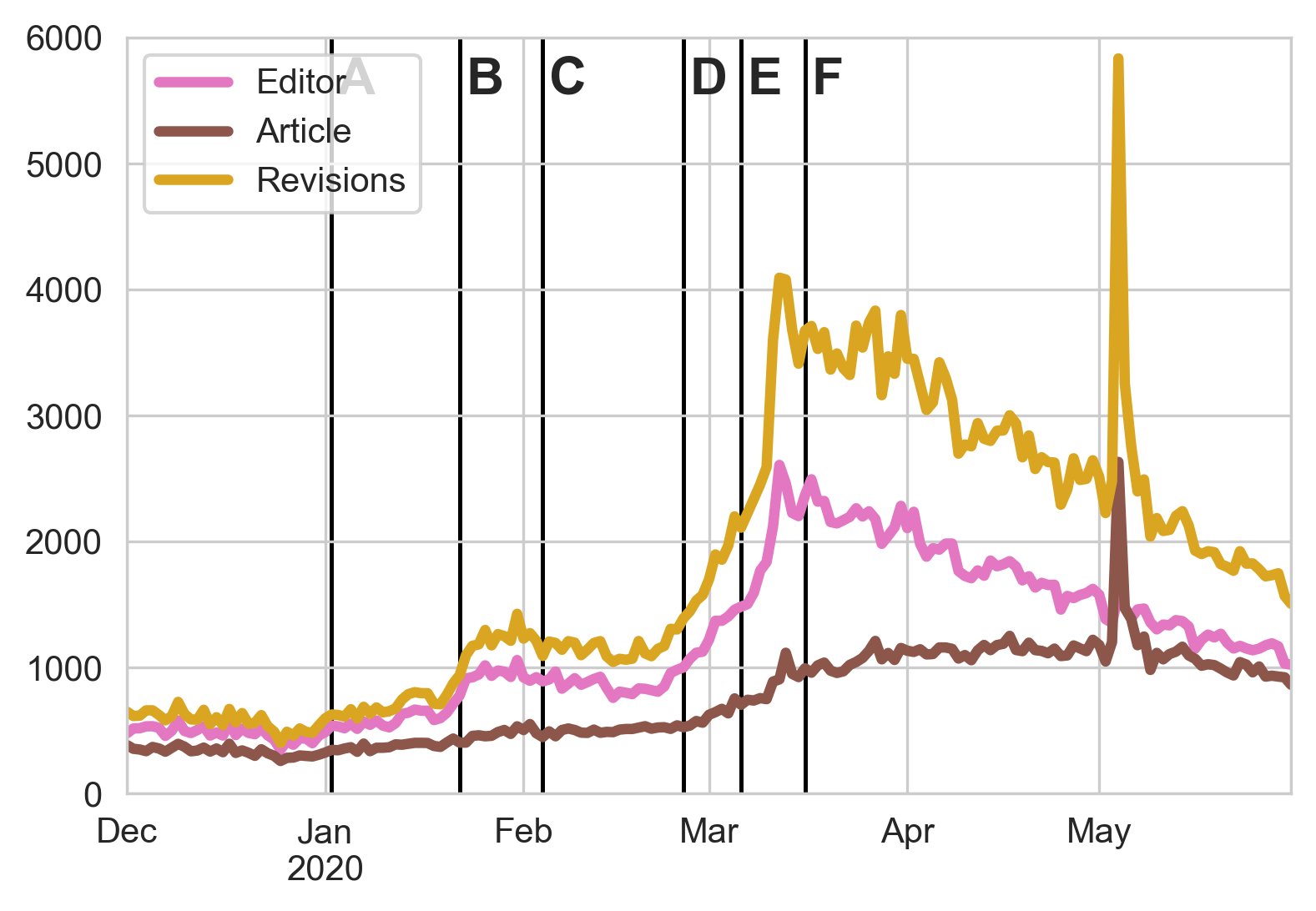}
        \caption{Size of the ``snapshot'' collaboration networks.
        }
        \label{fig:g_snapshot_counts}
    \end{minipage}\hfill
    \begin{minipage}{.475\textwidth}
        \centering
        \includegraphics[width=\textwidth]{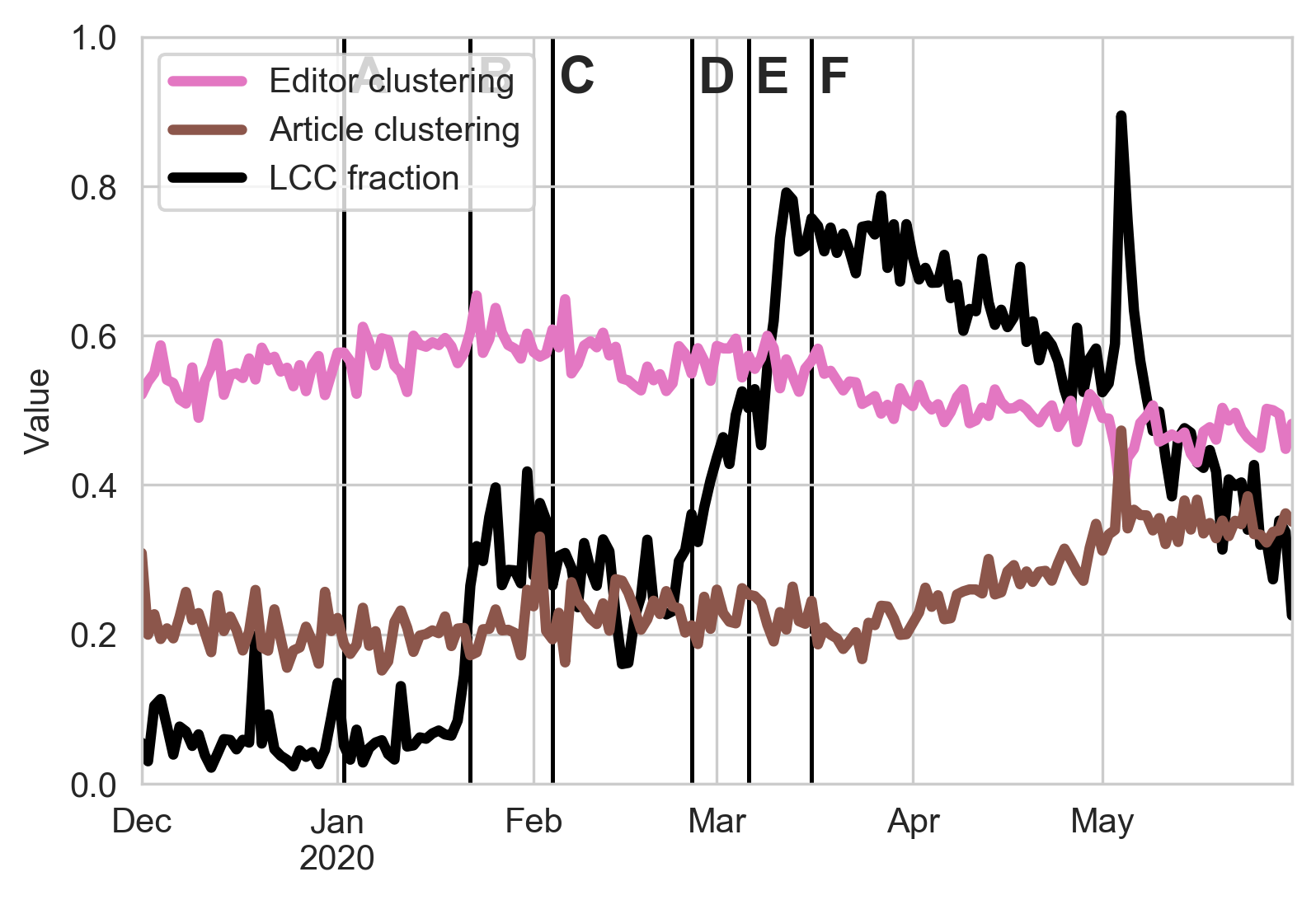}
        \caption{Changes in the clustering coefficients and LCC fraction in the ``snapshot'' collaboration networks.
        }
        \label{fig:g_snapshot_clustering_lcc}
    \end{minipage}
\end{figure}

Collaboration networks capture the structure of editors making contributions to articles. These bipartite graphs\footnote{Also known as two-mode or affiliation networks.} are unique because two distinct types of nodes exist and relationships can only exist between different node types~\cite{borgatti_network_1997}. In other words, editors cannot revise editors and articles cannot revise articles; only editors can revise articles. Bipartite graphs require distinct algorithms for analyzing their structure in light of this constraint~\cite{latapy_basic_2008}; we use the \texttt{bipartite} class within Python's \texttt{networkx} library~\cite{hagberg_exploring_2008}. The dynamics of Wikipedia's collaboration networks can be captured with two different strategies. The first strategy is a \textit{cumulative collaboration network} capturing \textit{all} the revision activity between editors and articles leading up to a given date. The second strategy is a \textit{snapshot collaboration network} capturing \textit{only} the revision activity between editors and articles on a given date. Because even significant changes in activity can be drowned out by the magnitude of previously accumulated behavior in the cumulative strategy, we visualize daily changes using the snapshot strategy in Figures~\ref{fig:g_snapshot_counts} and~\ref{fig:g_snapshot_clustering_lcc}.

Figure~\ref{fig:cuml_collab_g_distributions} captures the distributions of connectivity and activity in the cumulative collaboration network. We employ logarithmically-binned values in combination with double-logged axes to identify structure in the ``tails'' of the distribution~\cite{milojevic_binning_2010,virkar_powerlaw_2014}. There are ``long-tailed'' distributions in all three facets: most editors contribute to one article, most articles have few editors, and and most editors make a single revision (upper left of each facet) while some editors edit hundreds of articles, some articles have thousands of contributing editors, and some editors revise an article hundreds of times (lower right of each facet). While ``long-tailed'' distributions are ubiquitous in many complex systems like online social behavior~\cite{johnson_emergence_2014,mislove_measurement_2007}, these distributions are not necessarily generated by mechanisms like preferential attachment~\cite{andriani_perspectivegaussian_2009,broido_scalefree_2019,clauset_powerlaw_2009,mitzenmacher_brief_2004,milojevic_modes_2010}. We emphasize the truncated tails in all three of these distributions are deviations from true power laws and imply an over-representation of high-activity users, pages, and revisions.

Figure~\ref{fig:g_snapshot_counts} captures the size of the daily snapshot collaboration networks. The activity here mirrors the dynamics seen with revision (Figures~\ref{fig:daily_revisions_seed} and~\ref{fig:daily_revisions_all}) and pageview (Figure~\ref{fig:daily_pageviews_seed} and~\ref{fig:daily_pageviews_all}) activity: an initial rise in late January with coverage related to the initial outbreak in Wuhan, a rapid increase from late February through early March when major outbreaks are reported in Europe and the United States, and a decline in activity from the middle of March to the end of May as interest wanes. There is a significant spike of collaboration activity on May 4 when the primary seed article was renamed from ``2019--20 coronavirus pandemic'' to ``COVID-19 pandemic'' and required updating thousands of articles.

Figure~\ref{fig:g_snapshot_clustering_lcc} captures the ``cohesion'' of the daily snapshot collaboration networks using two types of features: average bipartite clustering coefficients~\cite{opsahl_triadic_2013,latapy_basic_2008} and fraction of nodes in the largest connected component (LCC fraction). The editor clustering coefficient captures the extent to which editors share articles in common and the article clustering captures the extent to which articles share editors in common. The clustering coefficient values are relatively stable until mid-March when the production and consumption of information from these collaborations shifts. Editor clustering (pink line) declines slightly, suggesting editors are less likely to work together and instead focus their contributions on fewer articles. Article clustering (brown line) increases substantially, suggesting articles are more likely to share editors. These seemingly contradictory results can be reconciled by considering the dramatic changes in the size of the collaborations during this time: there are many more editors than articles, new editors joining and new articles being created simultaneously make it harder for all editors to contribute to many articles in common, but the contributions from a handful of editors working across multiple articles increases the article clustering. The increases in article re-engagement (Figure~\ref{fig:daily_user_reengagement}) as well as the increasing cross-article engagement among the seed articles (Figure~\ref{fig:daily_user_reengagement_seeds}) also reinforce this interpretation.

The LCC fraction captures the extent to which the collaboration is fragmented across small and isolated engagements or coheres into a large and well-connected effort. The LCC fraction (black line) mirrors the dynamics seen in Figure~\ref{fig:g_snapshot_counts} with a pre-outbreak low connectivity, Wuhan-era moderate connectivity, and a pandemic-era high connectivity phases. The LCC fraction peaks with the mid-March surge of interest with an influx of new editors and article creations and then again in early May with the renaming event and approximately 80\% of editors and articles are indirectly connected together into a giant component. There are three distinct ``steps'' in the LCC fraction over time. The first step is a ``pre-pandemic'' phase ending in late January when articles are edited in relative isolation from one another: editors on the ``World Health Organization'' article in late December are unlikely to also revise the ``2020 Democratic Party presidential primaries'' article on the same day, so these articles and their editors belong to separate components in this analysis. In other words, the low LCC fraction at this time is the ``shadow of the future'' capturing the behavior of articles that do not yet know they will become linked together in the future through this pandemic event. The second step is an ``early pandemic'' phase from late January to mid-February when some articles begin to be edited together. The ``shadow of the future'' still looms and the relatively low LCC fraction continues to capture the extent to which many articles that later become related to the pandemic are not yet aware and so remain disconnected from the emerging coronavirus collaboration. The third step is the ``peak pandemic'' phase from late February through the end of May with a high LCC fraction as hundreds of previously unrelated articles are now updated on the same day to reflect the effects of the pandemic. The spike in early May is an artifact of the renaming event.

\subsection{Hyperlinks}

\begin{figure}[t]
    \centering
    \includegraphics[width=.67\textwidth]{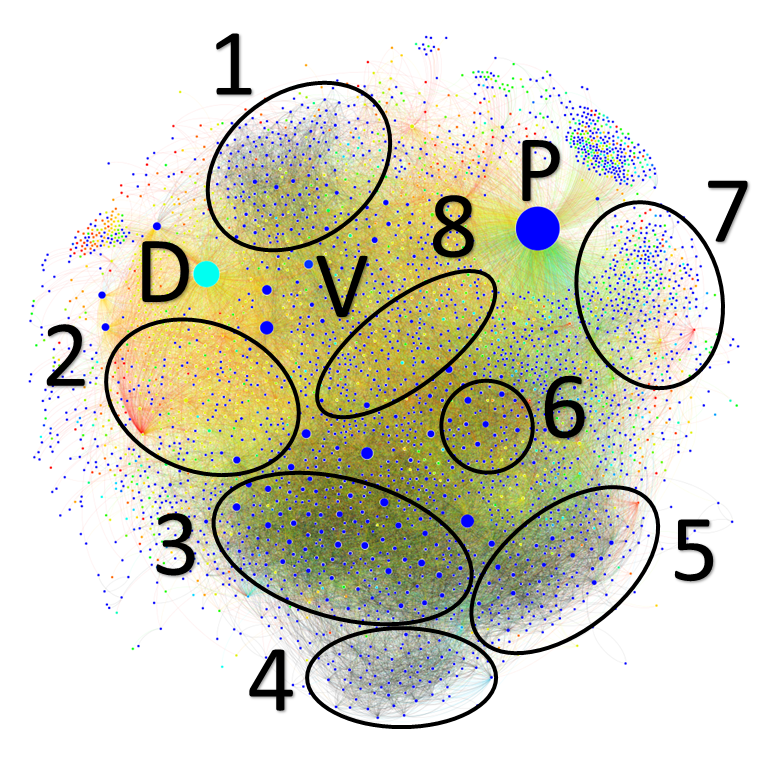}
    \caption{Spring layout visualization of the largest connected component in the cumulative hyperlink network. Nodes are sized by in-degree and colored by first appearance in the network (bluer earlier, redder newer). Edges are weighted by the number of weeks they appeared in the network and colored by first appearance in the network (bluer earlier, redder newer). The pandemic (P), disease (D), and virus (V) articles are annotated along with 8 specific clusters.}
    \label{fig:cuml_hyperlink_g}
\end{figure}

The hyperlink network captures the links from the body of an article to other Wikipedia articles. We only look at the hyperlinks within the combined set of category-related and neighbor-related articles. We purposely exclude links within templates and infoboxes because these are often repeated across articles (causing artificially dense networks) and more accurately captures the semantics of each article's content. Because some articles link to out-of-date (\textit{e.g.}, ``2019-20 Wuhan outbreak'') or non-standard article names that are then redirected to the canonical article title~\cite{hill_consider_2014}, we resolve each revision's outlinks to the canonical article title. Finally, we also retrieved the historical content of each article in the sample on a weekly frequency to analyze how the creation of new pages and links affected the structure of the hyperlink network over time.

Figure~\ref{fig:cuml_hyperlink_g} is a spring layout visualization of the ``cumulative'' hyperlink network that locates nodes with similar connections closer together. Nodes and edges are colored by the date of creation, with bluer colors corresponding to parts of the network that already existed or were introduced earlier and redder colors corresponding the parts of the network that were introduced later. The article about the pandemic event (P, approximately 1 o'clock), disease (D, approximately 10 o'clock), and virus (V, also 10 o'clock) are annotated. The sampling strategy ensures that this network is relatively densely-connected but there are several sub-communities with particularly dense connections annotated 1 to 5 counter-clockwise:
\begin{description}
    \item[Group 1.] Medical topics like ``Infection'', ``Pneumonia'', and ``Coronavirus''.
    \item[Group 2.] Locality-specific pandemic articles like ``COVID-19 pandemic in Italy''.
    \item[Group 3.] Geographic and historical articles like ``Italy'' and ``World War II''.
    \item[Group 4.] Air transportation articles like ``American Airlines'' and ``Boeing 737''.
    \item[Group 5.] Chinese social articles like ``Community Party of China'' and ``Pinyin''.
    \item[Group 6.] News and social media articles like ``The New York Times'' and ``Twitter''.
    \item[Group 7.] Impacted events like ``2020 Kentucky Derby'' and ``2020 U.S. presidential election''.
    \item[Group 8.] Institutions like ``University of Oxford'' and ``World Health Organization''.
\end{description}
\noindent Other topical categories are also present in the network but outside coherent sub-communities: biographies of political leaders, scientists, and victims; releases and events cancelled due to the pandemic; universities and companies impacted by and/or responding to the pandemic. The hyperlinks within well-established sub-communities like groups 3--6 have little sign of change: articles like the ``United Kingdom'' and ``World War II'' and the links between them existed before the pandemic began. The pandemic article was the first of the core articles created and the rainbow of hued links around it reflect the growth in the content of the article over this time window: older (bue and green) links going towards articles in Groups 3 and 5 and newer links (yellow and red) going towards articles in groups 1 and 2. The disease article was created four weeks later and is deeply embedded between groups and 2, reflecting the similarity in linking behavior 

\begin{figure}[t]
    \centering
    \includegraphics[width=\textwidth]{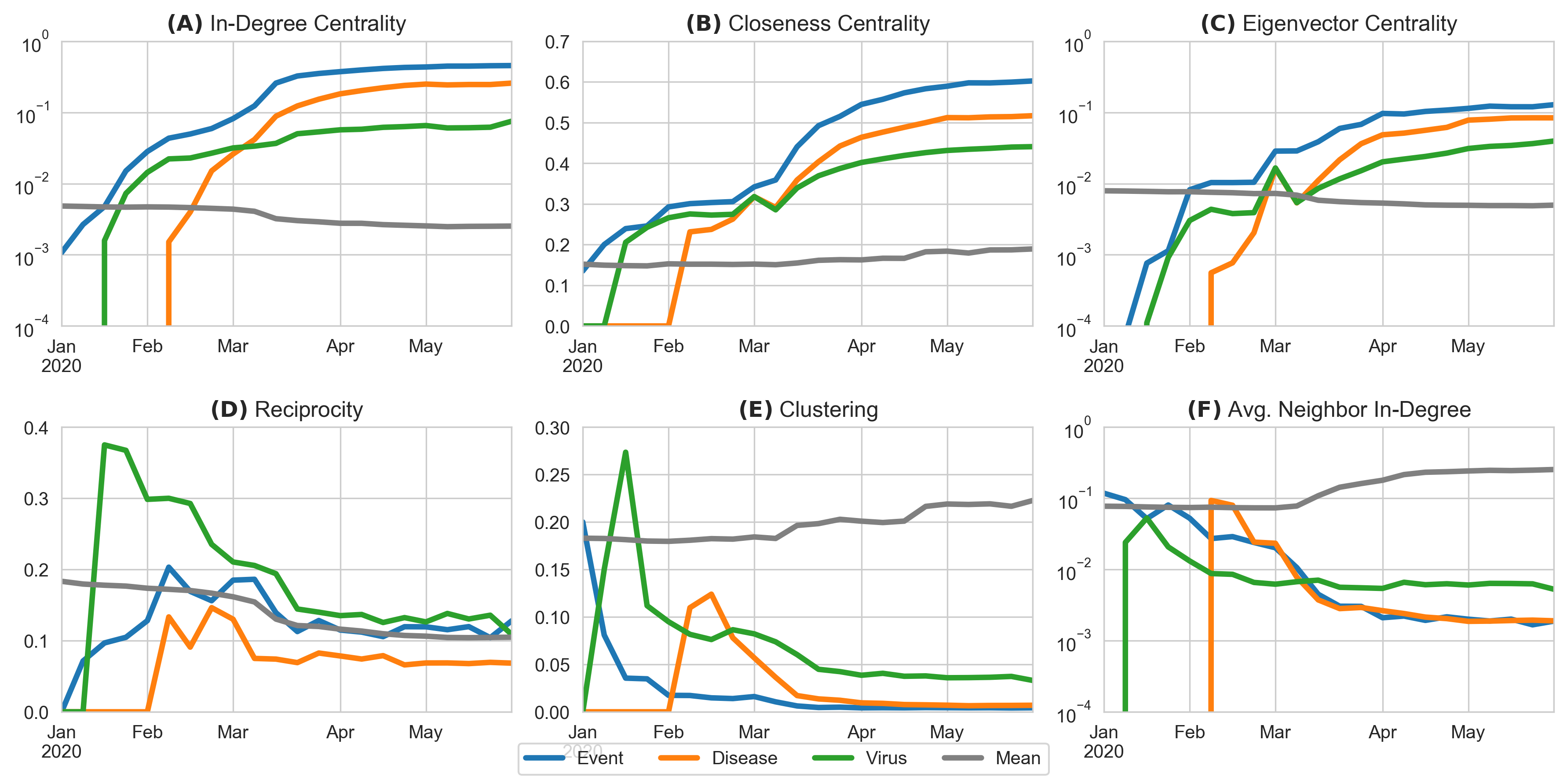}
    \caption{Changes in the in-degree centrality (A), closeness centrality (B), eigenvector centrality (C), reciprocity (D), clustering (E), and average in-degree neighbor's in-degree (F) of the pandemic (blue), disease (orange), virus (green), and mean (grey) articles within the weekly hyperlink networks.}
    \label{fig:hyperlink_weekly_seed_changes}
\end{figure}

Because of the difficulty parsing network visualization ``hairballs'', we analyze how the pandemic event, disease, and virus articles' positions in these networks changed over time. Figure~\ref{fig:hyperlink_weekly_seed_changes} plots six different features and the values for the three seed articles compared to the average values of all articles in the hyperlink network. Summarizing the findings below: the seed articles have many connections and became increasingly prominent in the hyperlink network, but many of these connections are from newly-created but poorly-connected articles. 


\begin{description}[leftmargin=1em]
    \item[In-degree centrality] (Figure~\ref{fig:hyperlink_weekly_seed_changes}A). In-degree centrality captures the number of other articles that link to an article. We report values normalized by the size of the network to capture the fraction of other nodes in the network connecting to the three seed articles. Unsurprisingly, all three seed articles' in-degree centrality grows significantly from less than 1\% of the nodes in the network connecting to them to 45\% of articles connecting to them by the end of May as other articles update their content and include hyperlinks to these articles. The average in-degree centrality for the rest of the network declines from March through May as hundreds of new articles are created (Figure~\ref{fig:daily_page_creations}) but receive few links from other articles in the network
    \item[Closeness centrality] (Figure~\ref{fig:hyperlink_weekly_seed_changes}B). Closeness centrality captures the average shortest path length an article has to every other article in the network. Lower closeness centrality implies greater ``distance'' or more other articles that need to be passed through to reach the rest of the network and higher closeness centrality implies less ``distance'' and more articles can be reached directly or through few other articles. All three seed articles' closeness centrality increases substantially from late February through the end of May as these articles link to and are linked from other articles in the network.
    \item[Eigenvector centrality] (Figure~\ref{fig:hyperlink_weekly_seed_changes}C). Eigenvector centrality captures the ``prestige'' of an article having connections from other ``prestigous'' articles. The seed articles' eigenvector centrality also increases significantly over the average article's eigenvector centrality starting in late February with the second revision surge. Receiving links from high-profile articles like ``United States'' or ``World Health Organization'' increased the eigenvector centrality of the seed articles.
    \item[Reciprocity] (Figure~\ref{fig:hyperlink_weekly_seed_changes}D). Node-level reciprocity is the ratio of hyperlinks pointing both to and from an article to the total hyperlinks around an article. Greater reciprocity implies a stronger relationship between articles as both articles link to each other. The reciprocity of the seed articles grows quickly tracking the initial growth of article size (Figure~\ref{fig:daily_size_all}), stabilizes through late March, then declines as the burst of newly-created national and local-level pandemic articles (Figure~\ref{fig:daily_page_creations}) are expanded with many more links coming into the seed articles (Figure~\ref{fig:hyperlink_weekly_seed_changes}A) but these seed articles not reciprocating by linking to minor articles like ``\href{https://en.wikipedia.org/wiki/COVID-19_pandemic_in_Easter_Island}{COVID-19 pandemic in Easter Island}''. The average reciprocity in the whole hyperlink network declines around the same time for similar reasons and the reciprocity for the seed articles is similar to the rest of the network.
    \item[Clustering] (Figure~\ref{fig:hyperlink_weekly_seed_changes}E). Clustering captures the fraction of an article's hyperlink neighbors that are connected to each other. Greater clustering implies an article is embedded with a cohesive sub-community while lower clustering implies an article's neighbors are poorly connected or the article is bridging between otherwise poorly connected sub-communties. The clustering for all three seed articles rises briefly in the earliest weeks following creation then falls dramatically as the articles become more prominent and surrounded by diverse topics that do not link to each other. The clustering for these seed articles is much less than the average value in the rest of the network, again reflecting the prominence of these articles in the network in connecting different topics together.
    \item[Average neighbor in-degree] (Figure~\ref{fig:hyperlink_weekly_seed_changes}F). The average neighbor in-degree is a measure of how well-connected the articles linking to an article are themselves. An article with few links from other articles (low in-degree) but those articles being highly-linked from other articles would have a high average neighbor in-degree. The values for the seed articles peak immediately after their creation and decrease significantly before stabilizing in March and April. This is a result of new child articles linking to the pandemic event, disease, and virus articles even if these articles receive few links from any other articles in the network. 
\end{description}




\section{Discussion}\label{sec:discussion}

How did Wikipedia's coverage of the coronavirus pandemic evolve? We explored this research question through three levels of analysis: the dynamics of articles' revision activity and content, the contribution behavior of editors, and the structure of collaboration and hyperlink networks. This analysis is an in-depth case study using quantitative methods to describe the emergent collaborations around a historic global disruption in nature. The individual results in the previous sections illuminate how information production and consumption on these articles changed profoundly and rapidly, contributing editors increased their engagement and production by shifting their attention, and networks of collaborators and content coalesce in the aftermath of events. 

These individual results provide overlapping, cumulative, and emergent evidence of more general themes enabling these high-tempo collaborations. We adopt approaches from qualitative research like open coding to identify emergent themes common across the quantitative results from each level of analysis. This combination of an in-depth case study structured as multiple levels of analysis, application of mixed quantitative methods across these levels, and inductive analysis of their results is a kind of ``quantitative portrait.'' Much like pointillist art aggregates individual points into coherent structures, a quantitative portrait aggregates simple data into intermediary descriptive results that then coalesce into more coherent themes. These themes also reveal strategies for engaging in quantitative portraiture like consistent annotation, derivative features, re-classifying sub-groups, and retroactive sampling. We summarize the themes in our portrait in Table~\ref{tab:findings} and describe how we arrived at these themes and their implications for doing quantitative portraiture in more detail below.

\begin{table}[t]
    \footnotesize
  \centering
  \begin{tabular}{p{0.8\textwidth}}
  \toprule
       {\bf Theme 1}: Event-driven activity and derivative features. \\
       {\bf Theme 2}: Core-periphery spillovers and expanding samples. \\
       {\bf Theme 3}: Contradictory user engagement and reclassification. \\
       {\bf Theme 4}: Shadow of the future and retrospective sampling. \\
    \bottomrule
  \end{tabular} \\ \vspace{1em}
  \begin{tabular}{ccp{8cm}cccc}
    \toprule
     & \textbf{Figure} & \textbf{Description} & \textbf{T1} & \textbf{T2} & \textbf{T3} & \textbf{T4} \\ \midrule
     \multirow{12}{*}{\STAB{\rotatebox[origin=c]{90}{\textbf{Article}}}}
     & \ref{fig:daily_revisions_seed} & Revision peaks follow external events but slow down since April & \cmark & \cmark & \cmark & \\
     & \ref{fig:daily_revisions_all} & Wide-spread and intense revision activity across articles & \cmark & \cmark & \cmark & \\
     & \ref{fig:editing_time_of_day} & Revision activity has consistent dirurnal cycles & & & \cmark & \cmark \\
     & \ref{fig:daily_edit_lag_all} & Revision persistence only minutes during most acute editing & & \cmark & \cmark  & \\
     & \ref{fig:daily_diff_all} & Revision size varies in response to intensity and events & & \cmark & \cmark \\
     
     & \ref{fig:daily_page_creations} & Surge of geographic child article creation to manage coverage & & & \cmark & \\
     & \ref{fig:daily_size_all} & Rapid growth in length of articles, surpassing high-quality neighbors & & & \cmark & \cmark \\
     & \ref{fig:genealogy} & Seed pages serve as parents of newly created pages & & \cmark & \cmark & \\
     & \ref{fig:weekly_seed_content_changes} & Substantial differences in content of articles week-over-week & & & & \cmark \\
     & \ref{fig:daily_pageviews_seed} & Double peak in attention, category overtakes neighbor  &\cmark & \cmark & \cmark&  \\
     & \ref{fig:daily_pageviews_all} & Millions of daily pageviews to pandemic-related articles in March & \cmark & \cmark & \cmark &  \\ \midrule
    
    \multirow{8}{*}{\STAB{\rotatebox[origin=c]{90}{\textbf{User}}}} 
     & \ref{fig:daily_cuml_users_seed} & Growth in number of collaborators tracks revision activity & & & \cmark & \\
     & \ref{fig:daily_user_reengagement} & High levels of editor retention day-over-day, week, and ever & & \cmark &  \\
     & \ref{fig:daily_user_reengagement_seeds} & Strong flows of editors between seed articles & & & \cmark & \\
     & \ref{fig:editing_intensity} & High and low-tempo contribution behaviors consistent over time & & & \cmark &  \\
     & \ref{fig:cohort_daily_activity} & Pre-pandemic editors remain the most active & \cmark & \cmark & \cmark & \\
     & \ref{fig:daily_account_age_seeds} & Ages of collaborators regress to the mean & \cmark & & \cmark & \cmark \\ 
     & \ref{fig:daily_edit_count_fraction_seeds} & Earnest novices replaced by grizzled veterans & & & \cmark & \cmark \\
     & \ref{fig:seed_users_pagetype_prepost} & Contribution behavior is disrupted after first seed edit & & \cmark & \cmark & \cmark \\
     \midrule
     
    \multirow{5}{*}{\STAB{\rotatebox[origin=c]{90}{\textbf{Network}}}} 
     & \ref{fig:cuml_collab_g_distributions} & Strong skews in activity across levels & & \cmark & \cmark & \\
     & \ref{fig:g_snapshot_counts} & Size of collaborations tracks revisions & \cmark & \cmark & & \cmark  \\
     & \ref{fig:g_snapshot_clustering_lcc} & Increasing coherence of collaboration & \cmark & & \cmark & \cmark \\
     & \ref{fig:cuml_hyperlink_g} & Clusters emerge among the category- and neighbor-related pages & & \cmark & & \cmark \\
     & \ref{fig:hyperlink_weekly_seed_changes} & Seed articles' network positions become prominent & \cmark & \cmark & & \cmark \\ 
    \bottomrule
  \end{tabular}
  \caption{Summary of findings from individual figures with thematic mapping (T1, T2, T3).}
  \label{tab:findings}
\end{table} 

\subsection{Theme 1: Event-driven activity and derivative features}
The shocks of significant external events influence information production and consumption behaviors that ripple through our results. Significant changes in the scale, tempo, and similarity of activity happened in the immediate aftermath of announcements as people sought, created, and updated information. Much of the activity is centered on two similar events: the announcement of quarantines, first in China in late January and then in Europe and the United States in the middle of March. These events drove surges of editing and pageview activity across our sample of thousands of articles when examined in the aggregate (Figures~\ref{fig:daily_revisions_seed}, \ref{fig:daily_revisions_all}, \ref{fig:daily_pageviews_seed}, \ref{fig:daily_pageviews_all}). These twin events oriented our interpretations of subsequent analyses---as seen in our repeated use of vertical annotations in time series---focusing on more abstract features. 

Annotation practices across diverse descriptive results enabled us to identify anomalies when the behavior of articles, editors, or networks either ``resisted'' the expected disruptions from external events or activity bursted or trends changed in the absence of any salient event. The absence of bursts and shifts in trends were both anomalies and invitations for ``zooming in'' to the underlying data to understand the mechanisms at play through both close interpretations of the underlying digital trace data~\cite{geiger_trace_2011} as well as exploration of related features. This could take the form of comparison to other time series for fingerprints of similar anomalies, manually inspecting archived revisions on Wikipedia itself to understand the contexts, or developing derivative aggregations or features to interrogate the assumptions of the first analysis. One example of this type of the last form of triangulation---analyzing derivative features---was testing whether the surprisingly high re-engagement statistics were robust across different sampling windows (Figures~\ref{fig:daily_user_reengagement} and~\ref{fig:daily_user_reengagement_seeds}) and finding similar trends. Grounded in the hypothesis that major events should structure the activity and behavior of these collaborations, consistent annotation of those events in our results enabled us to navigate findings across levels of analysis. 



\subsection{Theme 2: Core-periphery spillovers and expanding samples}
The multiple events unfolding across time as well as the proliferation of child articles about the regional, national, and sub-national effects of the pandemic created an unusually large ``periphery'' of articles surrounding the seed articles about a global pandemic, the disease, and the virus. Previous analyses of Wikipedia's breaking news collaborations as well as crisis informatics hashtag-centered sampling both bake in assumptions that privilege artifacts around singular event. The interactions afforded by a platform and behaviors queriable via an API contribute significantly to this privileging of ``core events'' while ignoring the effects of the event on a broader ecosystem of related content and behavior. Incorporating peripheral articles---category-related and neighbor-related articles---alongside our analysis of core ``seed'' articles revealed similarities and differences meriting further investigation.

Our sampling frame enabled to explore how the bursts of activity on the core articles were not isolated but also ``spilled over'' into the rest of the information ecosystem: editors made revisions across articles causing them to connect as networks and readers consumed information across articles causing their pageviews to spike in tandem. ``Zooming out'' from the core to the periphery enables us to see how distinct but similar articles behave differently following the same event or the profound scale of activity in the seed articles are yet still dwarfed by behavior unfolding in parallel around them. Grounded in the hypothesis that different types of similarity should reveal different kinds of behavior, our inclusive sampling approach identified two discrete sets of related articles that became a reliable basis for comparison with each other as well as with articles in the core (Figures~\ref{fig:daily_revisions_all}, \ref{fig:daily_size_all}, \ref{fig:daily_pageviews_all}). The similarities or differences between these distinct sets of core and peripheral articles enabled us to navigate findings across levels of analysis.



\subsection{Theme 3: Contradictory user engagement and reclassification} 
Because editors are the agents of so much of the change seen in the data, it is natural to ask who they are but the results paint a complex and conflicting picture. The tempo of editing articles can become so rapid that content is overwritten within minutes (Figure~\ref{fig:daily_edit_lag_all}) yet editors also do not deviate significantly from their diurnal patterns  (Figures~\ref{fig:editing_time_of_day} and~\ref{fig:editing_intensity}). A majority of editors make one or two contributions (Figure~\ref{fig:cuml_collab_g_distributions}) but more than 60\% of editors on any given day already contributed before (Figure~\ref{fig:daily_user_reengagement}). These contradictions arise from problems like Simpson's paradox that invite inquiry into how aggregations might be biased by different behaviors of unobserved sub-groups~\cite{alipourfard_using_2018,lerman_computational_2018}. 

In asking about unobserved sub-groups, we can also begin to reflect on the factors that bind users together into sub-groups and why those relationships might be hard to observe. For our questions around user engagement, we artificially stratified users by a seemingly arbitrary feature like temporal co-occurrence of first edits and yet recovered new perspectives (Figure~\ref{fig:cohort_daily_activity}). Grounded in the experience that conflicting evidence can be the product of confounded behavior, the results from these reclassifications of existing data clarify some of the observed contradictions: engagement can remain high with participation so strongly skewed to low-effort edits because there is a sub-population of committed editors whose commitment predates the collaboration and sustains it over time. All data analysis involves some form of coarsening, aggregation, and generalization but these introduce biases and leads to complexities. By revisiting our assumptions for aggregation, we opened new lines of inquiry to help in triangulating across levels of analysis and navigating conflicting results.

\subsection{Theme 4: Shadow of the future and retrospective sampling}
In game theory, the ``shadow of the future'' that explains how we behave differently when we expect to have interactions in the future. In the case of the prisoner's dilemma, you cooperate \textit{now} so you are not punished for defecting in the \textit{future}. Our sampling strategy is an interesting form of time travel in that we take what we know about the present (\textit{e.g.}, editors and articles linked together in the aftermath of a deadly global pandemic) and return to their past to effectively watch them get hit by a proverbial bus. But our units of analysis---articles and editors---are only in our sample because historic events they know nothing about at one time will eventually draw them together. 

Traditional methods of social inquiry do not have the luxury of high-resolution archival data that allow researchers to reconstruct the state of the world \textit{before} a disruption so they can track the behavior of people \textit{through} it. Analyses like the changes in content (Figure~\ref{fig:weekly_seed_content_changes}), cross-article re-engagements (Figure~\ref{fig:daily_user_reengagement_seeds}), or changes in network structure (Figure~\ref{fig:g_snapshot_clustering_lcc} and \ref{fig:hyperlink_weekly_seed_changes}) allow us to revisit these past events and to trace what people did at the time in the face of ambiguity and even danger. Editors have many first contributions, but some significantly change their behavior for weeks or months (Table~\ref{tab:behavioral_change} and Figure~\ref{fig:seed_users_pagetype_prepost}), which they do not know at the time. This retroactive sampling of users' behavior before their actions has remarkable (and ironic) parallels to epidemiologists' contact tracing. We started with a strategy for expanding the sample along a dimension of topical similarity, but temporal dimensions such as users' contributions before engaging in a collaboration or reconstructing a high-fidelity previous state of the world bring in other kinds of similarities and dependencies that can be analyzed. Of course, reliable access to these kinds of high-resolution archival data requires increasingly precious data infrastructures~\cite{boyd_critical_2012,Freelon_ComputationalResearchPostAPI_2018,Perriam_DigitalmethodspostAPI_2019}.


\subsection{Limitations and future work}
As with so much Wikipedia research, these analyses and findings use only data and behavior from the English Wikipedia. Because different language editions of Wikipedia exhibit significant differences in content and behavior~\cite{bao_omnipedia_2012,hecht_tower_2010}, our findings about how the English Wikipedia responds to crises may not generalize. As noted in the analyses above, page protection is a ubiquitous confounder in our analysis by severely limiting the kinds of users who can participate. But Wikipedia's page protections are an under-appreciated empirical setting for studying content moderation strategies: in the absence of market demands for engagements or subscriptions, Wikipedia can employ blunter strategies to cope with bad behavior. While the quantitative portraiture method we have outlined here has a greater commitment to induction and triangulation than most quantitative methods admit, it nevertheless involves significant abstractions of social behavior into aggregated counts~\cite{howison_validity_2011}. Qualitative methods like trace and virtual ethnography, interviews, and discourse analysis are still sorely needed to fully apprehend the dynamics animating the timing of changes to articles, the motivations of editors who self-select into these bizarre collaborations, and the values in play around debates about what kinds of content to include. 

\section{Conclusion}
Wikipedia has a dogged capacity to respond to current events over two decades. In the face of a historic global pandemic, enormous collaborations have self-organized to document information about the events of the pandemic, evidence about the disease, and the details about the virus as well as hundreds of more specific articles. Given the enormous space of issues, we performed a ``quantitative portrait'' using a variety of quantitative methods at multiple levels of analysis to understand the features of this remarkable collaboration. Using two dozen descriptive results, we identify four themes cutting across them and the implications these have for combining quantitative methods. 

\begin{acks}
Wikimedia Foundation for the data.
\end{acks}

\clearpage
\bibliographystyle{ACM-Reference-Format}
\bibliography{bibliography}

\end{document}